\documentclass[useAMS,usenatbib,referee]{mn2e}

\usepackage{graphicx}
\def\wes{CXOU\,J1647-4552\,}

\title[X-ray spectra from magnetar candidates]{
X-ray spectra from magnetar candidates. I. Monte Carlo
simulations in the non-relativistic regime
} \author[L. Nobili, R. Turolla and
S. Zane]{L. Nobili$^{1}$\thanks{E-mail: nobili@pd.infn.it (LN);
turolla@pd.infn.it (RT); sz@mssl.ucl.ac.uk (SZ)}, R. Turolla$^{1, 2}$ and
S. Zane$^{2}$\\ $^{1}$Department of Physics, University of Padova, via
Marzolo 8, 35131 Padova, Italy\\ $^{2}$Mullard Space Science Laboratory,
University College London, Holmbury St. Mary, Dorking, Surrey, RH5 6NT,
UK}

\begin{document}

\date{}

\pagerange{\pageref{firstpage}--\pageref{lastpage}} \pubyear{2007}

\maketitle

\label{firstpage}

\begin{abstract}

The anomalous X-ray pulsars and soft $\gamma$-repeaters are peculiar 
high-energy sources believed to host a magnetar, an ultra-magnetized 
neutron star with surface magnetic field in the PetaGauss range. Their 
persistent, soft X-ray emission exhibit a two component spectrum, usually 
modeled by the superposition of a blackbody and a power-law tail.  It has 
been suggested that the $\sim 1$--10 keV spectrum of AXPs/SGRs forms as 
the thermal photons emitted by the cooling star surface traverse the 
magnetosphere. Magnetar magnetospheres are, in fact, likely different from 
those of ordinary radio-pulsars, since the external magnetic field may 
acquire a toroidal component as a consequence of the deformation of the 
star crust induced by the super-strong interior field. In a twisted 
magnetosphere, the supporting currents can provide a large optical depth 
to resonant cyclotron scattering. The thermal spectrum emitted by the star 
surface will be then distorted because primary photons gain energy in the 
repeated scatterings with the flowing charges, and this may provide a 
natural explanation for the observed spectra. In this paper we present 3D 
Monte Carlo simulations of photon propagation in a twisted magnetosphere. 
Our model is based on a simplified treatment of the charge carriers 
velocity distribution which however accounts for the particle collective 
motion, in addition to the thermal one. Present treatment is restricted to 
conservative (Thomson) scattering in the electron rest frame. The code, 
nonetheless, is completely general and inclusion of the relativistic QED 
resonant cross section, which is required in the modeling of the hard 
($\sim 20$--200 keV) spectral tails observed in the magnetar candidates, 
is under way. The properties of emerging spectra have been assessed under 
different conditions, by exploring the model parameter space, including 
effects arising from the viewing geometry. Monte Carlo runs have been 
collected into a spectral archive which has been then implemented in the 
X-ray fitting package XSPEC. Two tabulated XSPEC spectral models, with and 
without viewing angles, have been produced and applied to the 0.1--10 keV 
{\em XMM-Newton} EPIC-pn spectrum of the AXP \wes.

\end{abstract}

\begin{keywords}
Radiation mechanisms: non-thermal -- stars: neutron -- X-rays: stars.
\end{keywords}

\section{Introduction}
\label{intro}

Over the last few years, increasing observational evidence has gathered in favour of the existence of
``magnetars'', i.e. neutron stars (NSs) endowed with an ultra-strong magnetic field ($B \approx
10^{14}-10^{15}$~G), much higher than the critical threshold at which quantum electro-dynamical (QED) effects
become important ($B_{crit} \sim 4.4 \times 10^{13}$~G). The existence of these objects has been first proposed
in the early '90s by \cite{dt92} and \cite{td93}, who suggested that, soon after
the core collapse following the supernova explosion, convective motions can strongly amplify the seed magnetic field
via helical dynamo action. The magnetar model,
initially developed to describe the phenomenology of the so-called soft $\gamma$-ray repeaters (SGRs), namely
the emission of strong bursts, the fast spin period evolution and the persistent X-ray luminosity, is currently
believed to successfully reproduce the properties of another class of peculiar NSs, the anomalous X-ray pulsars
(AXPs). Although alternative models, invoking accretion from a fossil disc, are not completely ruled out
by observations as yet \cite[see e.g.][]{vp95,cha00,ros00}, the recent detection of SGR-like bursts from five AXPs
\citep{gavrill,kaspi,woods05,kaspi4u,cxo}
has strengthened the connection between the two groups and pushed forward the interpretation of AXPs as
magnetars.

Both classes of sources, SGRs and AXPs, are characterized by
spin periods in a narrow range (5--12 s), a typical persistent
X-ray luminosity of $\approx 10^{34}$--10$^{36}\ {\rm erg\, s}^{-1}$, no
evidence for Doppler shifts in the light curve, lack of bright
optical companions (favouring an interpretation in terms of isolated objects),  and a
spin-down in the range 10$^{-13}$--10$^{-10}\ {\rm s\, s}^{-1}$. In particular,
the magnetar scenario appears promising in providing an alternative mechanism
(namely the ultra-strong magnetic field) to power
their high X-ray luminosity, which can not be otherwise explained in terms of more
conventional processes, as accretion from a binary companion or injection of rotational energy in the pulsar
wind/magnetosphere. Besides, measurements of period and period
derivative,  assuming that spin-down is associated to magneto-dipolar
losses, are strongly suggestive of the presence of an ultra-strong magnetic field,
$B > B_{crit}$.

The soft X-ray spectra of AXPs are generally well described by a two component model, consisting  of a blackbody
with $kT\sim$ 0.4--0.5\,keV, and a power-law with photon index $2\la\Gamma \la 4$ \citep[e.g.][and references
therein]{woodsrew}. In some cases, SGRs spectra have been fit with a single power-law component, but recent deep
observations showed that, also for these sources, a blackbody component is often required
\citep{sandro05a,sandro06}. Despite the fact that the blackbody plus power-law  model has been routinely applied
to magnetar candidate spectra for several years, attempts to provide a physical interpretation for these
two components have just begun.

Recently it has been proposed that this phenomenological spectral model mimics a situation in which soft seed
photons emerging, for instance,  from the neutron star surface are boosted to higher energies by efficient
resonant cyclotron scattering (RCS) from magnetospheric charged particles, leading to the formation of a
power-law high-energy tail. The basic idea has been discussed by \cite{tlk02} (TLK in the following), who
suggested that a possible difference between SGRs/AXPs and standard radio-pulsars is that in the former the
internal magnetic field is highly twisted, up to $\sim 10$ times the external dipole. Stresses imparted to the
star crust by the strong toroidal component of the internal magnetic field cause the crust to deform. This
produces, in turn, a displacement of the footpoints of the external magnetic field lines with the net result
that, at intervals,  the external (initially dipolar) field may acquire a toroidal component, i.e. it may twist
up as well. Twisted magnetospheres are threaded by currents, substantially in excess of the Goldreich-Julian
current. As shown by TLK, charge carriers may provide large optical depth to resonant cyclotron scattering so
that soft (thermal) photons produced at the star surface gain energy through repeated collisions with the moving
charges. Since the electron distribution is spatially extended, and the resonant cross section depends on the
{\it local} value of $B$, it is expected that repeated scatterings lead to the formation of a high-energy tail,
instead of a narrow cyclotron line. At least qualitatively, this scenario may also explain the correlation
between spectral hardening, luminosity and increase in bursting/glitching activity that has been recently
discovered in the long-term evolution of a few sources \citep{sandro05a,rea05,rea07}.

The problem of computing the X-ray spectrum emerging from twisted magnetospheres
has been previously tackled using a simplified one-dimensional approach by \cite{lg06},
and a systematic application to X-ray data has been presented by \cite{nanda2} (see also \citealt{nanda1}).
More recently, 3-D Monte Carlo calculations have been presented
by \cite{ft07}, although these spectra have never been applied
to fit X-ray observations. Both these investigations
treat resonant cyclotron scattering in the non-relativistic regime, and
neglect electron recoil, i.e. use the resonant cross section in the particle frame in
the (magnetic) Thomson limit.

Interestingly, thanks to {\it INTEGRAL\/}, it has been recently
found that AXPs and SGRs exhibit very hard high-energy tails
($\Gamma\sim1$) which can extend up to $\sim
200$~keV \citep{kuiper,denhartog,rev,mere05,molkov,dg06}.
This discovery come somewhat as a surprise, being the persistent spectra
of these sources below $\sim 10$ keV rather soft, and changed our
view of magnetars, suggesting that their luminosity might well be
dominated by the hard, rather than soft, X-ray component.
The origin of such high energy tails is presently unclear, but,
again, most of the scenarios proposed so far invoke emission from
magnetospheric particles. Quite recently, \cite{tb05} discussed how soft gamma-rays may be
produced in a twisted magnetosphere, suggesting two different
mechanisms: either thermal bremsstrahlung emission from the surface
region heated by returning currents, or synchrotron emission from
pairs created higher up ($\sim$ 100 km) in the magnetosphere. However, an
alternative possibility is that the high-energy tails are again created
by resonant magnetic Compton up-scattering of soft X-ray photons. In order to boost
efficiently soft photons up to a few hundred keVs, scattering must
occur off a non-thermal population of relativistic electrons (or pairs), possibly located close
to the stellar surface \citep{baring}. A quantitative calculation of the expected spectra, which
necessarily requires a correct description of relativistic effects,
has not been put forward as yet.

In this paper, the first in a series devoted to investigate the X-/soft $\gamma$-ray persistent spectrum of
magnetar candidates, we lay out the physical bases of our model and present a Monte Carlo
code which is used to follow the  spectral modifications as the soft seed photons get progressively
up-scattered in the magnetosphere of an ultra-magnetized neutron star. Our present goal is to test,
by direct comparison with observations, if RCS spectra are capable of accounting for the observed properties
of the soft X-ray emission ($\la 10$~keV) of SGRs/AXPs. To this end, we adopt a
non-relativistic (Thomson) description for the scattering process.
However, the numerical scheme is completely general and is explicitly designed
to incorporate the fully QED cross sections and to deal with more complex magnetic configurations.
The former, together with an application to the hard X-ray tails detected by {\it
INTEGRAL}, will be the scope of forthcoming papers (Nobili, Turolla \& Zane in preparation).
In many respects the present investigation follows an approach similar to
that of \cite{ft07}, and we will refer to this paper for some useful
expressions. The two treatments, however, differ in a number of ways.
In particular, the present model includes the angular
and frequency dependence of seed photons in a more general way and a different prescription for the
current velocity distributions. Differences and similarities between the two methods will be
discussed along the paper, when relevant.

The paper is organized as follows. In \S\ref{model} we lay out and
scrutinize the physical bases of our model. The Monte Carlo
method and its coding is described in \S\ref{mc}, while in \S\ref{results} we present
the computed spectra and discuss their properties.  The implementation in
XSPEC of our model is described in \S\ref{arch} where also a preliminary
fit is reported. Discussion follows in \S\ref{disc}.

\section{The model}
\label{model} In this section we discuss in some detail the main
ingredients used in our computation of the soft ($\sim 0.1$--10 keV)
X-ray spectrum emitted by magnetar candidates.

\subsection{External magnetic field geometry}
\label{twistfield}

The first ingredient of our computation is a prescription for the magnetic
field geometry. Monte Carlo techniques are suitable for
handling complicated 3D configurations, and our code is
completely general from this point of view. However, for the sake of
simplicity, in this paper we restrict ourselves to the axially symmetric
twisted magnetosphere configurations studied by TLK, in which case the
(numerical) solution of the magnetostatic, force-free equilibrium is straightforward.
Accordingly, we report here only those expressions that
are needed to facilitate the reading of this paper, and refer to TLK
for all details.

The starting point is the force-free equation $\vec j\times\vec B=0$ where
$\vec j$ and $\vec B$ are the current and the external field respectively. Under
the assumption of axial symmetry, this equation can be written as
$\nabla\times\vec B=\alpha({\cal P})\vec B$ with ${\cal P}={\cal
P}(r,\theta)$ the flux parameter.
A major simplification
arises by
restricting to self-similar configurations, ${\cal P} = {\cal P}_0 r^{-p}
F(\cos \theta)$, in which case the problem reduces to the solution of a
second order eigenvalue differential equation for $F(\cos\theta)$, that can be solved
numerically for each value of the parameter $0\leq p \leq 1$. The latter univocally fixes
the magnetic configuration, a part for a scale factor $B_{pole}$ (see below).
The boundary conditions are chosen in such a way that the resulting axially-symmetric
configuration corresponds to a core-centered, twisted, dipolar field (see \S
\ref{disc} for a discussion).
The polar
components of the magnetic field are then (see again TLK for all details)

\begin{eqnarray}\label{btwist}
B_r & = & -\frac{B_{pole}}{2} \left(\frac{R_{NS}}{r}\right)^{2+p} \frac{dF}{d\cos\theta}\\
\nonumber
B_\theta & = & \frac{B_{pole}}{2} \left(\frac{R_{NS}}{r}\right)^{2+p} \frac{pF}{\sin\theta}\\
\nonumber
B_\phi & = & B_\theta \left [
\frac{C}{p \left ( 1 + p \right ) } \right ] ^{1/2} F^{1/p}\, ,
\end{eqnarray}
where the constant $C$ is an eigenvalue which depends on $p$ only,  $R_{NS}$ is the neutron
star radius and $B_{pole}$ is the value of the magnetic field at the pole. The net twist angle
is defined as

\begin{equation}\label{twistang}
\Delta \phi_{N-S} = \lim_{\theta_0 \to 0} 2\int_{\theta_0}^{\pi/2} \frac{B_\phi
}{B_\theta } \frac {d\theta}{\sin \theta}
\end{equation}
and is a function of the parameter $p$. As a consequence, either $p$ or $\Delta \phi_{N-S}$ can be used
to label each model in the sequence.

\subsection{Magnetospheric currents}
\label{currents}

Once the magnetic structure is known, in the force-free approximation the spatial
density of the magnetospheric particles is automatically fixed by

\begin{equation}\label{magcurr}
n_e ( \vec r, \beta ) = \frac{ p + 1 } {4 \pi e} \left ( \frac{B_\phi
}{B_\theta} \right ) \frac {B }{r\vert  \langle \beta \rangle\vert } \, ,
\end{equation}
where $\langle\beta\rangle$ is the average charge velocity (in units of $c$; see below). The above expression
gives the co-rotation charge density of the space charge-limited flow of ions and electrons from the NS surface,
that, due to the presence of closed loops in a twisted field, is much larger than the Goldreich-Julian density,
$n_{GJ}$. Moreover, it is important to note that, while a space charge-limited flow with $n = n_{GJ}$ requires
currents flowing in opposite directions from the two poles, for the case
at hand there is a well defined flow
direction which is the same from north to south. This breaks the symmetry between the two star hemispheres, and
implies that the observed spectrum will  be different when viewed from the north or the south pole. Clearly,
because of charge neutrality, the electron current must be balanced by ions flowing in the opposite direction.
However, ions are heavier, they are not lifted much in the
magnetosphere and tend to move closer to the star
surface. Photons may scatter off ions, but this is likely to  give rise at most to a narrow absorption feature
at the ion cyclotron energy (see TLK and \citealt{ft07}). For this reason, the ion current is not considered
here, together with pair creation, that can further complicate the relation between charge and current density
by introducing bidirectional flows  (see \S\ref{disc} for a discussion).
In a genuinely static 
twist ($\partial\vec B/\partial t=0$) the electric and magnetic fields are
orthogonal. This implies that the voltage drop between the footpoints of a field line vanishes since
$E_\parallel=0$, so that there is no force that can extract particles from the surface and lift them against
gravity thus initiating the current $\vec j_B=c\nabla\times\vec B/4\pi$ requested to support the twist. However,
as discussed in \cite{beth07}, once implanted, the twist has necessary to decay precisely to provide the
potential drop required to accelerate charges. A non-vanishing $E_\parallel$ is maintained by self-induction and
the twist evolution is regulated by the balance between the conduction current $j$ and $j_B$, 
$\partial E_\parallel/\partial t=4\pi(j_B-j)$. If $j<j_B$
the magnetosphere becomes charge starved and $E_\parallel$ grows at the expenses of the magnetic field, injecting
more charges into the magnetosphere. On the other hand, when $j>j_B$ the field decreases reducing the current.
The magnetosphere is then in dynamical (quasi)equilibrium with $j\sim j_B$
over a timescale $< t_{decay}$, where $t_{decay}\approx$ a few years is the twist decay time \citep{beth07}.

The second key ingredient is the velocity distribution of the
magnetospheric charges. This is a crucial and still largely unexplored issue
\cite[see however][and \S\ref{disc}]{beth07}. Nevertheless, in a
strong magnetic field the electron distribution is expected to be largely
anisotropic:
$e^-$ stream freely along the field lines, while they are confined in a set of
cylindrical Landau levels in the plane perpendicular to $\vec B$. In order
to mimic such scenario, we assume a
1-D Maxwellian distribution at a given temperature $T_e$, superimposed to a
bulk motion with velocity $v_{bulk}$, as measured in the stellar frame.
The (invariant) distribution function turns out to be

\begin{equation} \label{distribf}
\displaystyle{\frac{d n_e}{d (\gamma\beta) }} = \displaystyle{
\frac{n_e\, \exp{(-\gamma'/\Theta_e )} }{2 \, K_1(1/\Theta_e)}}
=  n_e f_e(\vec r , \gamma \beta)
\end{equation}
where $\gamma' = \gamma\gamma_{bulk} (1-\beta\beta_{bulk} )$, $\Theta_e = kT_e/m_e c^2$, $K_1 $ is the
modified Bessel Function of the first order and $f_e = \gamma^{-3} 
n_e^{-1} d 
n_e/d\beta$ is the momentum distribution function.
We consider $T_e$ and $\beta_{bulk}=v_{bulk}/c$ as free parameters in our model, and,
although this is definitely a simplification,
we assume that both do not depend on position.
This expression differs from that used by \cite{ft07} (their equation [19]) inasmuch
they do not include the effects of collective (bulk) velocity
(which is necessary to reproduce the current flow), but only those of
the $e^-$ local velocity distribution (either thermal, as in the present case, or non-thermal).
In other words, we assume that electrons move
isothermically along the field lines but, at the same time, they receive the
same boost from the electric field.
Even in the lack of
any detailed information about the charge accelerating mechanisms, we consider
our choice more realistic.

\subsection{Scattering cross sections}
\label{cross-sections}

Scattering off free electrons in the presence of a strong magnetic field
has been extensively treated in the literature. The non-relativistic
($B\ll B_{crit}$) expressions for the scattering cross sections in the
Thomson limit (i.e. neglecting electron recoil) were derived by
\cite{vent79} (see also \citealt{mes92}). The complete QED compton cross
sections have been presented by \cite{her79}, \cite{dh86}, and
\cite{hd91}. The scattering cross section depends on the incident photon
polarization state and, in general, it must be computed by summing over
the (infinite) virtual intermediate Landau states. Moreover, proper
account has to be made for the electron spin transition and for the
possibility that scattering leaves the electron in an arbitrary excited
state (Raman scattering). This leads to quite cumbersome expressions (see
e.g \citealt{hd91}), even if one restricts to the resonant part of the
completely differential cross section. On the other hand, under the
typical conditions expected in a twisted magnetosphere, soft photons
($\hbar\omega\sim 1$ keV) will undergo resonant scattering when
$\omega\sim\omega_B$ and this happens only where the field has decayed to a
value $B\sim 10^{11}\, {\rm G}\ll B_{crit}$. Electron recoil starts to be
important when the photon energy in the electron rest frame becomes
comparable to the electron rest energy. If $\gamma$ is the mean electron
Lorentz factor, this occurs at typical energies $\sim m_ec^2/\gamma$.
Assuming mildly relativistic particles, the previous limit implies that
conservative scattering should provide good accuracy up to photon energies
of some tens of keV. This, together with the fact that resonant
scattering occurs in regions where $B\ll B_{crit}$, makes the use of the
(much simpler) non-relativistic (Thomson) cross section adequate. We
anticipate here that, albeit supported by physical considerations, this
provides only a zeroth level description and a more thorough treatment
demands for the full QED cross section, as it is discussed in more detail
later on (see \S\ref{disc}).
A further simplification arises because, under the typical
conditions encountered in the magnetosphere, vacuum polarization dominates
over plasma effects. In this situation, the two (ordinary and
extraordinary) normal modes are linearly polarized.

Since radiative de-excitation occurs on a very short timescale, one can
safely assume that the electron is initially in the ground state. For a
particle initially at rest, the non-relativistic scattering cross sections
at resonance are easily derived from the general expression given e.g. by \cite{her79}
by performing the substitution
\begin{equation}
\displaystyle\frac{1}{(\omega-\omega_B)^2} \to  \frac{1}{(\omega-\omega_B)^2+\Gamma^2/4}
\end{equation}
where $\Gamma = (4e^2\omega_B^2)/(3m_ec^3)$ 
accounts for the finite transition lifetime
of the excited state \cite[e.g.][]{dv78, vent79}. 
Since in the present case it is $\hbar\omega_B\simeq\hbar\omega\sim
1$ keV, the resonance peak is so narrow and 
prominent that non-resonant contributions
to the cross section are negligible. One can therefore take the limit
\begin{equation}
\displaystyle\lim_{\Gamma\to 0}\frac{\Gamma}{(\omega-\omega_B)^2+\Gamma^2/4}=2\pi\delta(\omega-\omega_B)
\end{equation}
which results in
\begin{eqnarray} \label{difcross}
\nonumber
&\displaystyle\frac{d\sigma}{d\Omega'}\Big\vert_{1-1}&=\frac{3\pi
r_0c}{8}\delta(\omega-\omega_B)\cos^2\theta\cos^2\theta'\\
&\displaystyle\frac{d\sigma}{d\Omega'}\Big\vert_{1-2}&=\frac{3\pi
r_0c}{8}\delta(\omega-\omega_B)\cos^2\theta\\
\nonumber
&\displaystyle\frac{d\sigma}{d\Omega'}\Big\vert_{2-2}&=\frac{3\pi
r_0c}{8}\delta(\omega-\omega_B)\\ \nonumber
&\displaystyle\frac{d\sigma}{d\Omega'}\Big\vert_{2-1}&=\frac{3\pi
r_0c}{8}\delta(\omega-\omega_B)\cos^2\theta'
\end{eqnarray}
where
$\theta$ ($\theta'$) is the photon angle before (after) the
scattering and $r_0$ is the classical electron radius.
Here and in the following the index 1 (2) stands for the
ordinary (extraordinary) mode.

Upon normalization, the previous expressions give the probability that
an incident photon with polarization state $i$ and direction
$\theta$ is scattered at angle $\theta'$ with polarization state
$j$.  The total cross sections
for separated processes are easily computed by integrating the previous expressions over all outgoing
photon angles

\begin{eqnarray} \label{totcross}
\nonumber
&\displaystyle\sigma_{1-1}&=\int_{4\pi}d\Omega'\frac{d\sigma}{d\Omega'}\Big\vert_{1-1}=\frac{\pi
r_0c}{2}\delta(\omega-\omega_B)\cos^2\theta\\
&\displaystyle\sigma_{1-2}&=\int_{4\pi}d\Omega'\frac{d\sigma}{d\Omega'}\Big\vert_{1-2}=\frac{3\pi
r_0c}{2}\delta(\omega-\omega_B)\cos^2\theta\\
\nonumber
&\displaystyle\sigma_{2-2}&=\int_{4\pi}d\Omega'\frac{d\sigma}{d\Omega'}\Big\vert_{2-2}=\frac{3\pi
r_0c}{2}\delta(\omega-\omega_B)\\
\nonumber
&\displaystyle\sigma_{2-1}&=\int_{4\pi}d\Omega'\frac{d\sigma}{d\Omega'}\Big\vert_{2-1}=\frac{\pi
r_0c}{2}\delta(\omega-\omega_B)\, .
\end{eqnarray}
The total cross section for scattering of an incident ordinary (extraordinary)
photon is obtained by summing the first (second) pair of expressions in
equation (\ref{totcross}). Finally, in order to determine the photon direction
after scattering (i.e. the two angles $\theta'$, $\phi'$) in the Monte
Carlo code, the following integrals are required

\begin{eqnarray} \label{partcross}
\nonumber
&\displaystyle\frac{1}{\sigma_{i-j}}\int_{0}^{\phi'}\int_{0}^{\pi}d\Omega'\frac{d\sigma}{d\Omega'}\Big\vert_{i-j}&=
\frac12\phi'\\
&\displaystyle\frac{1}{\sigma_{1-1}}\int_{0}^{2\pi}\int_{0}^{\theta'}d\Omega'\frac{d\sigma}{d\Omega'}\Big\vert_{1-1}&=
\displaystyle\frac{1}{\sigma_{2-1}}
\int_{0}^{2\pi}\int_{0}^{\theta'}d\Omega'\frac{d\sigma}{d\Omega'}\Big\vert_{2-1}
=\frac{1}{2}(1-\cos^3\theta')\\
\nonumber
&\displaystyle\frac{1}{\sigma_{1-2}}\int_{0}^{2\pi}\int_{0}^{\theta'}d\Omega'\frac{d\sigma}{d\Omega'}\Big\vert_{1-2}&=
\displaystyle\frac{1}{\sigma_{2-2}}\int_{0}^{2\pi}\int_{0}^{\theta'}d\Omega'\frac{d\sigma}{d\Omega'}\Big\vert_{2-2}=
\frac{1}{2}(1-\cos\theta')
\end{eqnarray}

\subsection{Photon propagation in the magnetosphere}
\label{photons}

The scattering cross sections discussed in \S \ref{cross-sections} hold in
the electron rest frame (ERF). In particular, both the photon ($\omega$) and the cyclotron ($\omega_B$)
frequency entering expressions (\ref{difcross})--(\ref{partcross}) are evaluated in the ERF.
In the case of a charge moving with velocity $v=\beta c$ and Lorentz factor $\gamma$ with
respect to a frame attached to the star, the total cross sections (eq. [\ref{totcross}]) take the
form

\begin{eqnarray}
\label{totcrosslab}
&\displaystyle\sigma_{1-1}&=\frac{1}{3}\sigma_{1-2}=\frac{\pi^2
r_0c}{2}\delta(\omega-\omega_D)\cos^2\theta\\
\nonumber
&\displaystyle\sigma_{2-2}&= 3\sigma_{2-1} = \frac{3\pi^2
r_0c}{2}\delta(\omega-\omega_D)
\end{eqnarray}
where
\begin{equation}\label{omegad}
\omega_D=\frac{\omega_B}{\gamma(1-\beta\mu)}\, ,
\end{equation}
$\theta$ is the angle between the incident photon direction and the particle velocity as measured
in the ERF and $\mu$ is the cosine of the same angle but
measured in the stellar frame. The latter two quantities are related by the usual transformation
\begin{equation}\label{aberr}
\cos\theta=\frac{\mu-\beta}{1-\beta\mu}\, .
\end{equation}
Since particles are moving along $\vec B$, the magnetic field is unaffected by the Lorentz transformation,
and the value of $B$ as measured in the stellar frame can be used to compute the cyclotron frequency
$\omega_B$ in the ERF. It is worth stressing that $\omega$ in eqs. (\ref{totcrosslab})
is now the photon frequency in the stellar frame.

The scattering optical depth for a photon which travels a distance 
$d\ell$ in the magnetosphere is
\begin{equation}\label{tau}
d\tau_{ij}=d\ell\int_{\beta_{min}}^{\beta_{max}}d\beta n_e(\vec r) 
\gamma^3 (1-\beta\mu)
\sigma_{ij}(\omega,\vec r,\beta)f_e(\vec r, \gamma \beta)
\end{equation}
where the factor $1-\beta\mu$ appears because of the change of reference 
between the ERF
and the stellar frame, $n_e$ is the
(velocity integrated) particle density and $f_e$ 
is the (normalized) momentum distribution
as defined in equations (\ref{magcurr}) and (\ref{distribf}).

The indices $i$ and $j$ refer to the initial and final photon polarization states and $[\beta_{min},\,
\beta_{max}]$ is the charge velocity spread. As pointed out by \cite{ft07}, the integral in equation
(\ref{tau}) can be readily calculated by exploiting the $\delta$-function in the scattering cross section.
Denoting by

\begin{equation}\label{beta12}
\beta_{1,2}= \frac{1}{\mu^2+(\omega_B/\omega)^2}\left[\mu\pm\frac{\omega_B}{\omega}
\sqrt{(\omega_B/\omega)^2+\mu^2-1}\right]
\end{equation}
the two roots of the quadratic equation $\omega=\omega_D$,
the $\delta$-function in frequency can be transformed into a
$\delta$-function in velocity
\begin{eqnarray}\label{delta_beta}
&\delta(\omega-\omega_D)=&\frac{1}{\omega_B}\sum_{k=1,2}
\frac{(1-\mu\beta_k)^2}{\gamma_k\vert\mu-\beta_k\vert}\delta(\beta-\beta_k)=
\frac{\omega_B}{\omega^2}\sum_{k=1,2}
\frac{1}{\gamma_k^3\vert\mu-\beta_k\vert}\delta(\beta-\beta_k)\, .
\end{eqnarray}
Accordingly, the total scattering depth can be expressed as
\begin{eqnarray}\label{tautot_ord}
&d\tau_1=&d\tau_{1-1}+d\tau_{1-2}=
2\pi^2r_0c\frac{n_e\omega_B}{\omega^2}d\ell\sum_{k=1,2}
\frac{\vert\mu-\beta_k\vert}{ 
(1-\mu\beta_k)}f_e(\vec r, \gamma_k \beta_k)
\end{eqnarray}
and
\begin{eqnarray}\label{tautot_extr}
&d\tau_2=&d\tau_{2-2}+d\tau_{2-1}=
2\pi^2r_0c\frac{n_e\omega_B}{\omega^2}d\ell\sum_{k=1,2}
\frac{(1-\mu\beta_k)}{
\vert\mu-\beta_k\vert}f_e(\vec r,\gamma_k \beta_k)
\end{eqnarray}
for photons initially in the polarization state 1 and 2, respectively. 
These expressions
are analogous to those derived by \cite{ft07}, although their equation (33) seems to contain
an error. 
The spatial distribution of charged particles $n_e$ depends, 
in fact, on their 
average velocity, i.e. the speed at which charge carriers flow, and not on the velocity of the
single particle. For this reason $n_e$ is not evaluated at $\beta=\beta_k$, but at $\langle\beta\rangle$
which is, in general, a function of position (see also \S \ref{currents}).
Moreover, their equation (13) contains an unexpected factor $\omega_D$ in
place of $\omega_B$.
The reason for this is obscure since the ratio $\omega_B/B$ turns out
to be independent of both the magnetic field and photon energy.

Once the initial photon polarization, energy and direction have been fixed, equation (\ref{tautot_ord}),
or (\ref{tautot_extr}), is integrated along the photon path until a scattering occurs (see
\S \ref{scatdepth}). Although general relativistic effects are certainly important, here we restrict ourselves to
newtonian gravity and assume that photons move along straight lines between two successive scatterings. Proper
inclusion of null geodesics in a Schwarzschild space-time, albeit conceptually simple, turned out to be
computationally quite costly and we decided to dismiss it. As it is apparent from equation (\ref{delta_beta}),
resonant scattering may occur only when the roots $\beta_k$ are real, i.e. only
if $(\omega_B/\omega)^2+\mu^2-1\geq 0$. Since $\omega_B$ depends (through $B$) on position alone, at
every point in the magnetosphere the previous condition discriminates those pairs of photon energy
and angle for which scattering is possible \citep{ft07}. In case the particle velocity is always
of a given sign (charge carriers all positive or negative), only the roots $\beta_k$ with the same
sign are meaningful. If there exist two roots with the right sign (i.e. both are positive or negative),
the criterion for selecting onto which particle (the one with velocity $\beta_1$ or $\beta_2$) the photon
actually scatters is discussed in \S \ref{scatdepth}.

\subsection{Seed photon distribution}
\label{seed}

Primary photons are assumed to be emitted by the cooling surface
of the neutron star. Although, up to now, no detailed model for
surface emission from a magnetar has been presented, it seems
unlikely that the spatial and energy distribution of the
surface-emitted photons are the same as in ordinary cooling
neutron stars. In particular, being the surface heated
by returning currents (e.g. TLK), the surface temperature is
expected to be inhomogeneous (with the equatorial belt hotter than
the polar regions) and it is unclear if a standard (i.e. in
hydrostatic and radiative equilibrium) atmosphere can be present
on the top of a magnetar \cite[see however][]{guv06, guv07}.

On the wake of this, in order to keep our treatment as general as possible, we do not
prescribe an a priori surface temperature distribution (see \S \ref{photem}).
In the present version of the code, the initial energy distribution is taken
to be planckian for either ordinary and extraordinary photons, although other
spectral distributions can be easily accommodated. Different degrees of
polarization of the primary spectrum can be then obtained adding together, in
different proportions, ordinary and extraordinary blackbody photons.
Since the non-isotropic opacity of the stellar crust might convey
radiation in a preferred direction, we introduce a beaming parameter $b\ge 1$
such that the specific intensity at the star surface takes the form

\begin{equation}\label{seedinten}
n_\nu(\mu)\propto\mu^{b-1}\frac{\nu^2}{\exp({h\nu/kT})-1}
\end{equation}
where $\mu$ is the cosine of the angle between the initial photon direction and
the magnetic field. For $b=1$ the radiation is emitted isotropically
in the outward hemisphere.

\section{The Monte Carlo method}
\label{mc}

The code is structured into four main blocks, as outlined below. In the first
thermal photons are emitted from the stellar surface, in the second
the program evaluates the optical depth of the photon as it propagates through
the magnetosphere, while the third is finalized to solve the kinematics of the electron-photon
scattering. Finally, escaping photons are stored. Each block is briefly described in the following.

\subsection{Photon emission}
\label{photem}
Because of the intrinsic asymmetry of the
model, the observed spectrum depends on both the shape, and the (longitudinal) position
of the emitting region on the star surface, and the viewing direction.
Moreover, as mentioned above, the star surface temperature distribution
may not be isotropic. To account for these effects, the star surface is divided
into $N_\Theta\times N_\Phi$ zones by means of an equally spaced $\cos\Theta$ and $\Phi$ mesh, where
$\Theta$ and $\Phi$ are the magnetic colatitude and longitude.
This choice guarantees that all patches have the same area, so that the number of
emitted photons depends only on the
patch temperature (i.e. patches at the same
temperature emit the same number of photons).
A different temperature may be attached to each surface patch in
such a way to reproduce (up to the accuracy allowed by the finite
mesh resolution) any kind of thermal surface map.

Initially we fix the coordinates of an emitting patch and assign a value
for the polarization state $s$ of each seed photon, i.e. $s=1$
for the ordinary mode or $s=2$ for the extraordinary mode. All photons are emitted at
the patch centre $P$. Then, a photon is extracted at random from the distribution
(\ref{seedinten}). We assume that the initial photon angles are such that
the azimuth (as referred to $\vec B$ in $P$)  is uniformly
distributed while $\mu=\cos\theta_P$ is obtained solving the equation
$\mu = (U_P)^b $, where $ U_P$ is an uniform deviate and $b$ is the
beaming parameter introduced in eq. (\ref{seedinten}).
The coordinates of the emission point
and the initial momentum univocally determine the ray along which the photon moves.

Actually, after experiencing scattering(s), some photons will reach the star
surface again. Their number is fairly limited, since scattering typically occurs
at a distance $R_{sc}$ of a few stellar radii. The star disc, as seen from the last
scattering point, subtends a solid angle $\sim (R_{NS}/R_{sc})^2\la 0.1$, and this
is also an upper limit to the fraction of photons which are scattered back onto the
star surface. Numerical simulations show that the actual value is quite smaller,
$\la 1\%$. We assume that all photons impinging on the surface are absorbed (regardless
of their polarization state).

\subsection{Scattering depth}
\label{scatdepth}
In a Monte Carlo scheme the distance $\ell$ a photon of polarization state
$s$ travels between
two successive interactions (i.e. emission-scattering or scattering-scattering)
is estimated by integrating
the scattering depth $d\tau_s$ given by equations (\ref{tautot_ord}) and (\ref{tautot_extr})
until
\begin{equation} \label{integ_tau}
\tau_{s} = \int_0^\ell d\tau_s = -\ln U
\end{equation}
where $U$ is an uniform deviate.
Direct numerical evaluation of the integral (\ref{integ_tau}) proved, however, quite time consuming, and
we found more efficient and faster to perform a stepwise integration the differential equations
(\ref{tautot_ord}) and (\ref{tautot_extr})
using a fourth order Runge-Kutta method. Integration is terminated as soon as
the value of the optical depth exceeds $-\ln U$ and a linear interpolation between the last two steps
is used to determine with better accuracy the value of $\ell$  where $\tau_s = -\ln U$.

At each integration step we check if the photon still lies in the region of the $(\omega_b/\omega,\,\mu)$ plane
where resonant scattering is allowed, i.e. if $\omega_B^2/\omega^2+\mu^2-1\geq 0$ (see \S \ref{photons}).
When the previous inequality is found to hold no more, we further check if the photon trajectory
is bound to bring it back into the scattering permitted region or not.
This is achieved by computing numerically the tangent to the photon path [in the $(\omega_B/\omega,\,\mu)$ plane] where
$\omega_B^2/\omega^2+\mu^2-1\sim 0$ and checking if it lies in between the two limiting values
$(\mu\pm 1)/(\omega_B/\omega)$. If not, the photon is taken to
freely escape to infinity [see also figure 1 of
Fernandez \& Thompson (2007)]. The values of the energy and direction of the photon are then stored,
the program returns to step 1, and a new seed photon is emitted.

\subsection{The scattering process}
\label{scattering}
Assuming that equality $\tau_s=-\ln U$ is verified at some distance $\ell$ from the point of
the previous photon interaction, the kinematics of the scattering must be solved in order to obtain the new
direction and energy of the photon. This requires the knowledge of the velocity $\beta_k$ of the
resonant electron and the new photon polarization state. This is obtained by generating two new random numbers,
$U_1$ and $U_2$, and comparing them with the ratios of the corresponding cross sections. For a photon initially
in the ordinary polarization state ($s=1$), mode switching upon scattering
occurs if $ U_1 >
\sigma_{1-1}/(\sigma_{1-1}+\sigma_{1-2}) = 1/4$, while for an initially extraordinary photon ($s=2$) this
happens if $U_1 > \sigma_{2-2}/(\sigma_{2-2}+\sigma_{2-1})=3/4$. Similarly, the decision about onto which of the
two resonant electrons (assuming that both values of $\beta_k$ are acceptable) scattering actually occurs is
reached by comparing $U_2$ with the ratio $S_s(\beta_1)/[S_s(\beta_1) + S_s(\beta_2)]$, where $S_s(\beta_k)$
stands for each addendum in the sum at left hand side of equation (\ref{tautot_ord}) ([\ref{tautot_extr}]). If
$U_2<S_s(\beta_1)/[S_s(\beta_1) + S_s(\beta_2)]$, the scattering electron velocity is $\beta_1$, otherwise it is
$\beta_2$. At this stage, all parameters entering the differential cross section of the process are known.

Upon scattering with a moving charge, the momentum and energy of the photon are modified. Since the
cross section (\ref{partcross}) are defined in the ERF, the evaluation of the scattering angles $\theta'$ and
$\phi'$ requires a  Lorentz transformation from the stellar frame to the frame comoving with the resonant
electron $\beta_k$. For linearly polarized incoming light the distribution
of the azimuthal angle is isotropic,
so that $\phi' = 2\pi U_3$, where $U_3$ is an uniform deviate. Concerning the scattering angle, we note that
in the non relativistic case all quantities (\ref{partcross}) are proportional either to $1-\cos\theta'$ or to
$1-\cos^3\theta'$. Then, after drawing a new uniform deviate $U_4$, the scattering angle is given by
$\cos\theta'=2U_4-1$ or $\cos^3\theta'= 2U_4-1$, depending on the case.

The corresponding angles in the stellar frame and, hence the new photon direction, are obtained by means of
Lorentz transformations. In this frame the photon frequency is given by
\begin{equation}
\omega' = \gamma_k^2\omega\, (1-\beta_k\mu)(1+\beta_k\cos\theta')\, .
\end{equation}
Finally, once energy and momentum of the scattered photon are known the computation proceeds starting again from
point 3.2, integrating equations (\ref{tautot_ord}) or (\ref{tautot_extr}) along the new photon path.

\subsection{Photons storage}
\label{storage}
Escaping photons are collected on the ``sky at infinity'', i.e. on a spherical
surface located sufficiently farther out to see the star (and its magnetosphere)
as point-like. We introduce an angular grid $(\Theta_s,\, \Phi_s)$ which
divides the ``sky at infinity'' in a fixed number of patches, similarly
to what has been done for the stellar surface. When the escape condition (see
\S \ref{photons}) is met, the two angles $\Theta$ and $\Phi$ which characterize the
ray relative to the star centre are computed from the photon momentum and
the sky patch hit determined. Counts are stored in a three-dimensional array,
the first two indices of which label the sky patch while the third the photon
energy. This allows to analyze the
resulting spectra in different directions of observation when a large
number of events are processed. Each run involves $N_{tot}$ photons,
and is performed changing the initial polarization states $s$ and the
co-ordinates of the emitting patch. The resulting spectrum is obtained by
superposition of the various emitting patches.

\section{Results}
\label{results}

Our Monte Carlo code, written in FORTRAN90, proved to be efficient and
relatively fast. Despite the complexity of the whole procedure, we can
process about 7000 photons/s on a dual-core Xeon 2.8 GHz machine. The CPU
time for a typical production run (several million photons)
is 10--20 min. We stress that the result of each
run is a 3D array which
gives the number of counts at different positions on the sky
and at different energies (see \S\ref{storage}). Further manipulations
(e.g. to account for viewing angles, or to derive the pulse shape, see
below) are performed at the post-processing level by means of IDL scripts,
at negligible computational cost.
In the following subsections we discuss the general properties
of our spectral models.

\subsection{Spectra}
\label{spect}
In order to explore the role of the different parameters we computed a
set of spectra, by evolving $N_{patch} = 150,000$ photons for
$N_\Theta\times N_\Phi = 8 \times 4 =32$ surface patches (i.e. each model
has $N_{tot} = 4,800,000$ photons). We assume that the
star surface is at constant temperature, and that the seed radiation is
isotropic ($b = 1$, see \S~\ref{seed}) and completely polarized, either in
the ordinary or extraordinary mode. Furthermore, we treat the case of an aligned rotator, i.e.
the spin and magnetic axes coincide. Photons are collected onto a
$N_{\Theta_s} \times N_{\Phi_s} = 10 \times 10$ angular grid on the
sky, and in $N_E= 50$ energy bins in the range 0.1--100~keV.
The magnetic field has been fixed at $B=10^{14}$~G and the surface temperature at
$kT=0.5$~keV. The mean and the maximum number of scatterings per photon are in
the ranges $\sim 0.5$--2 and $\sim 10$--20, respectively, depending on the parameter values and on the
location of the emitting patch on the star surface.

\begin{figure*}
\hbox{
\includegraphics[width=84mm]{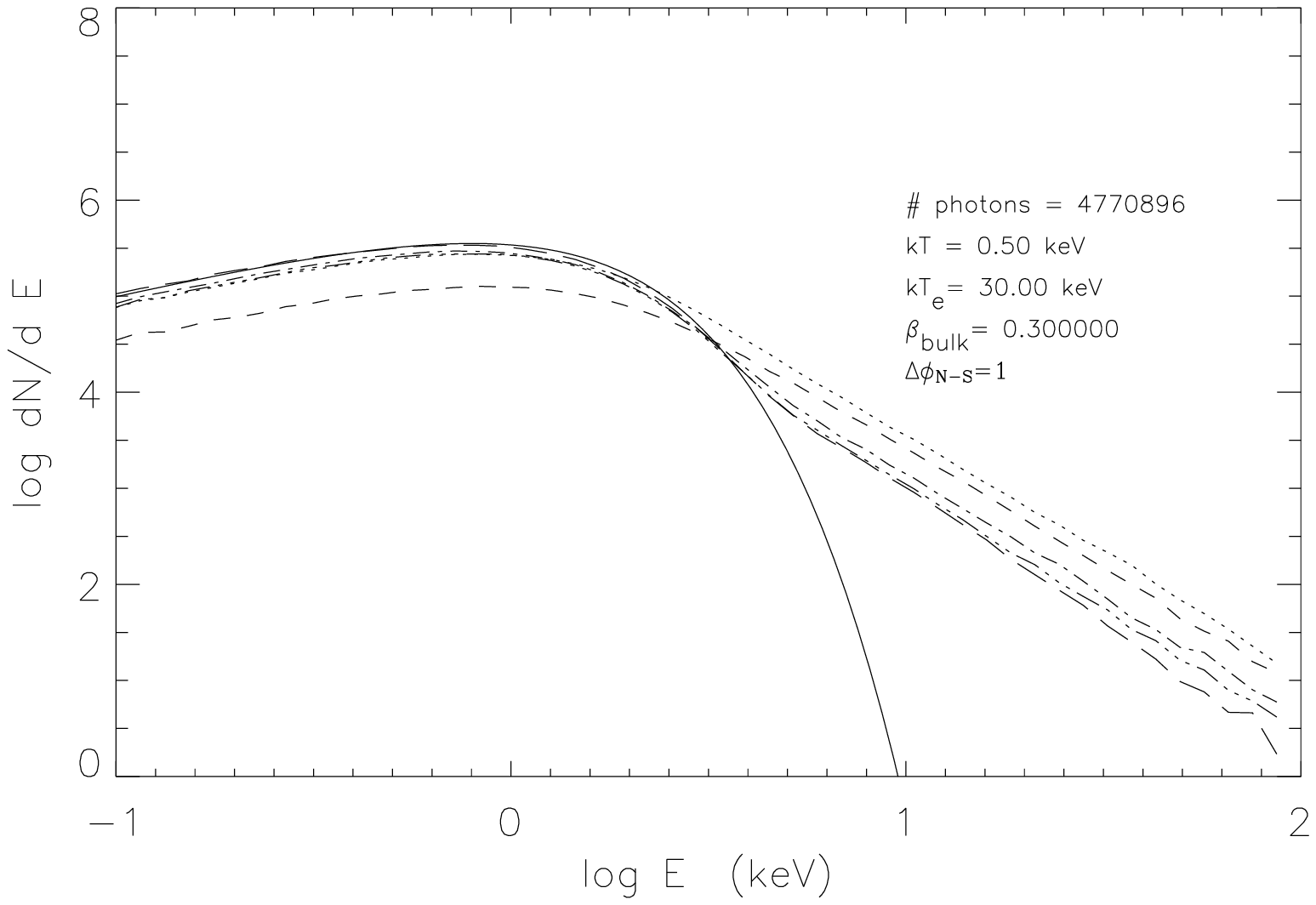}
\includegraphics[width=84mm]{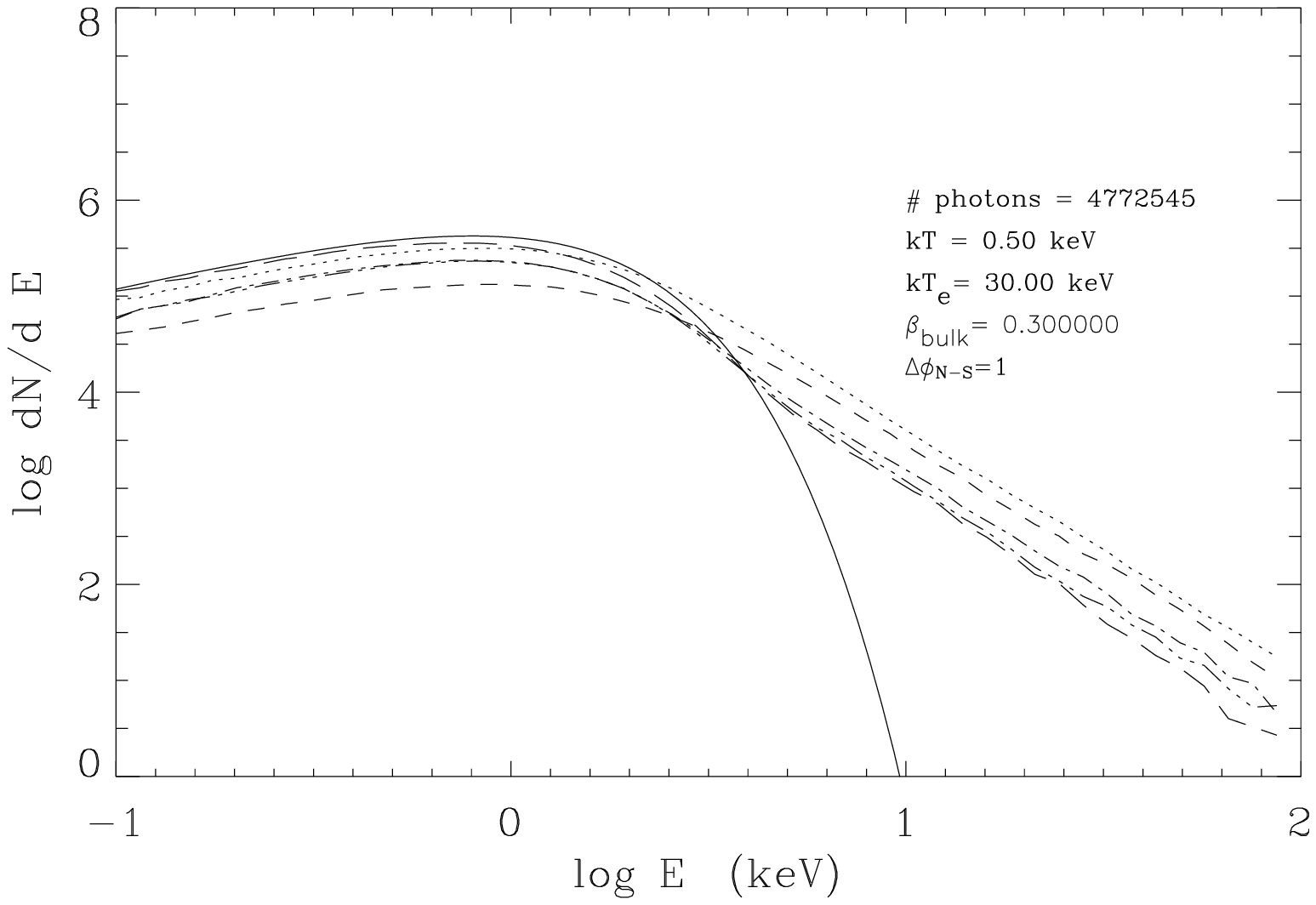}
}
\caption{Left: Computed spectra for $B = 10^{14}~G$, $kT  = 0.5$~keV, $kT_e =
30$~keV, $\beta_{bulk} = 0.3$, $\Delta \phi = 1$ and different values of
the sky colatitude
$\Theta_s$: $27^\circ$ (long dashed),
$64^\circ$  (dash-triple dotted), $90^\circ$
(dash-dotted), $116^\circ$ (short dashed) and
$153^\circ$ (dotted). The solid line represents the seed
blackbody and counts have been summed over $\Phi_s$.
Here seed photons are assumed to be completely polarized in the ordinary mode.
Right: Same, but for seed photons
completely polarized in the extraordinary mode.
\label{f1}
}
\end{figure*}

In Fig.~\ref{f1} we show the spectra, averaged over $\Phi_s$, as seen by
observers whose line-of-sight (LOS) is at different angles $\Theta_s$ with the star
spin axis\footnote{Note that
the total number of {\em collected} photons is usually lower than  $N_{tot}$ (4,800,000
in the present case) since a (small) fraction of photons reach infinity with an energy
outside our range of collection (i.e. 0.1--100~keV).}. The most salient
characteristic is the absence of symmetry between the north and the south
hemispheres: as $\Theta_s$ increases, spectra become more and more
comptonized. This reflects our choice for the electron velocity distribution,
which accounts for the charges bulk velocity, and currents
flow from the north to the south pole along the field
lines (of course the opposite choice for the current direction would
simply result in $\Theta_s \to 180^\circ - \Theta_s$). We found that the spectral
shape is almost insensitive to the seed photons polarization state (see Fig.~\ref{f1}).
This means that observations of the
phase averaged spectrum are not expected to provide useful insights into
the polarization degree of the surface emission (but see \S\ref{res_pol}).

\begin{figure*}
\hbox{
\includegraphics[width=84mm]{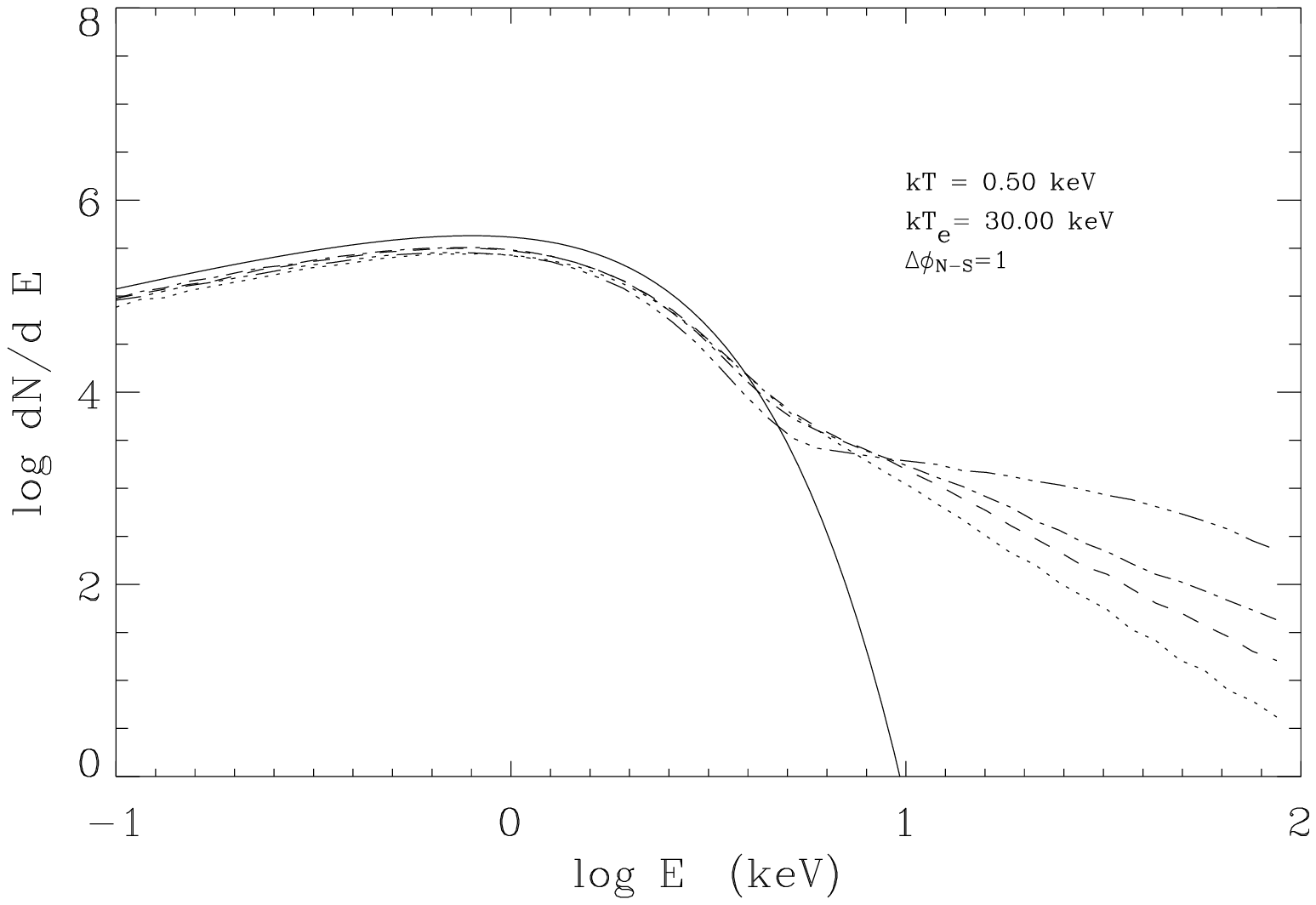}
\includegraphics[width=84mm]{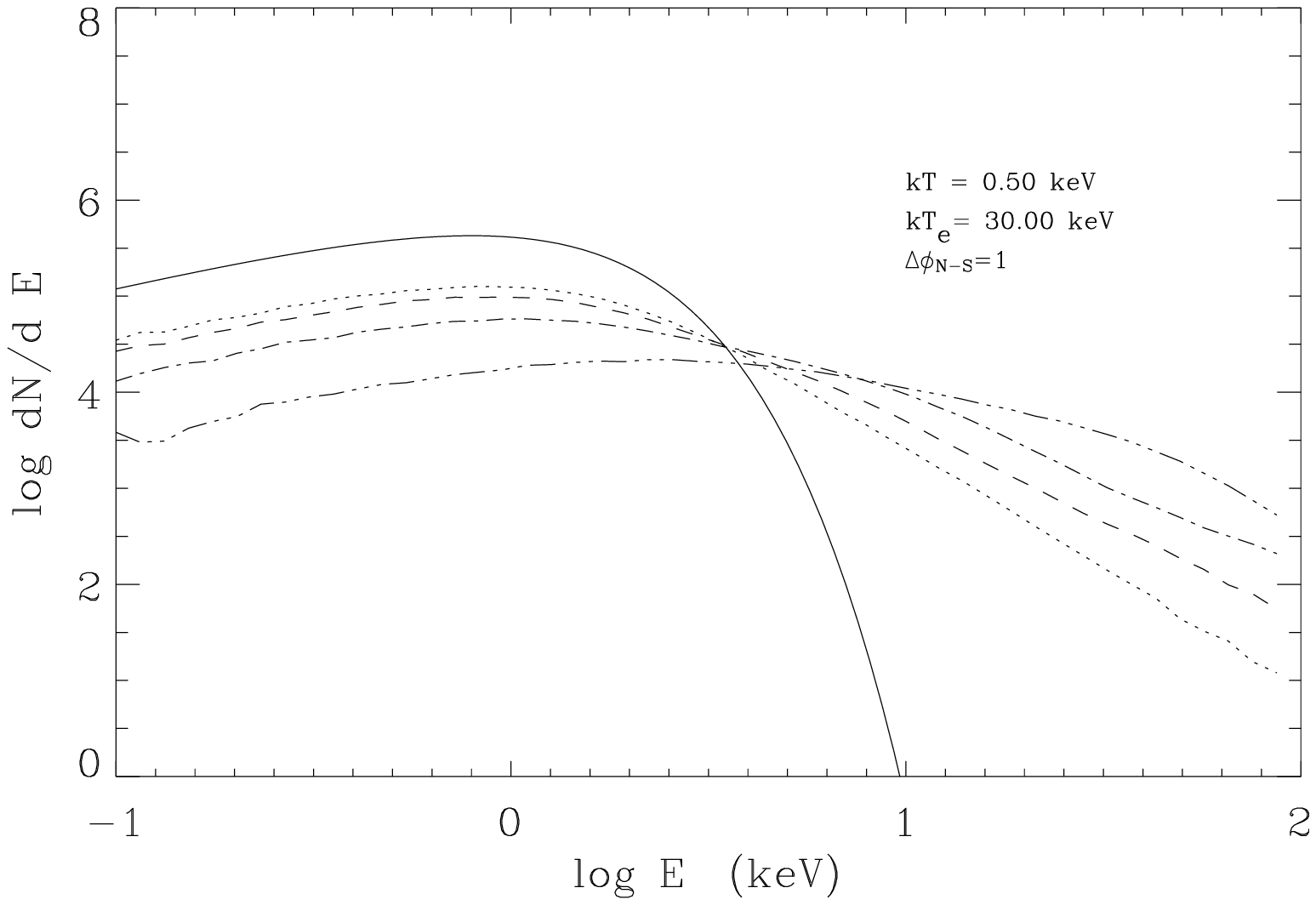}
}
\hbox{
\includegraphics[width=84mm]{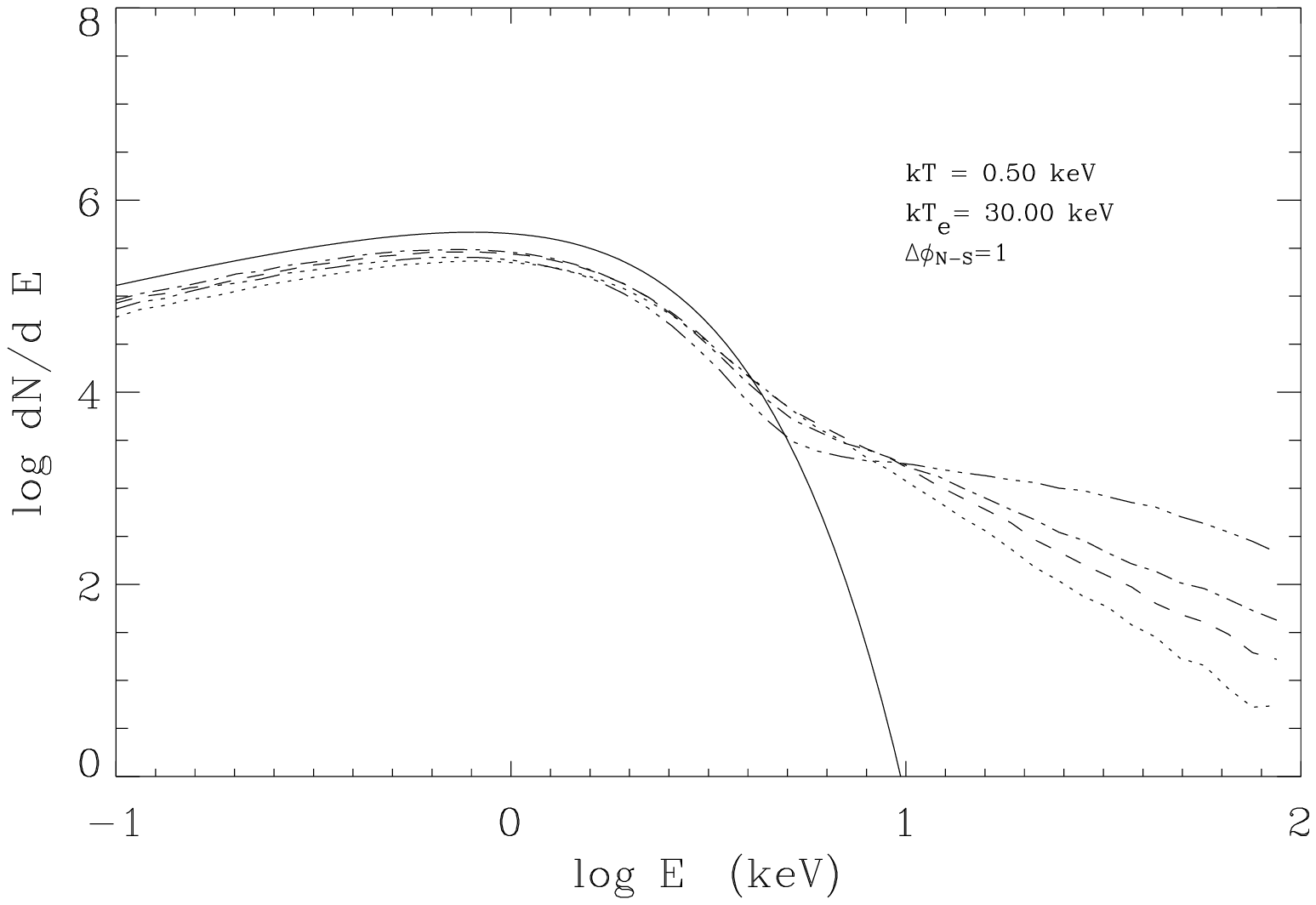}
\includegraphics[width=84mm]{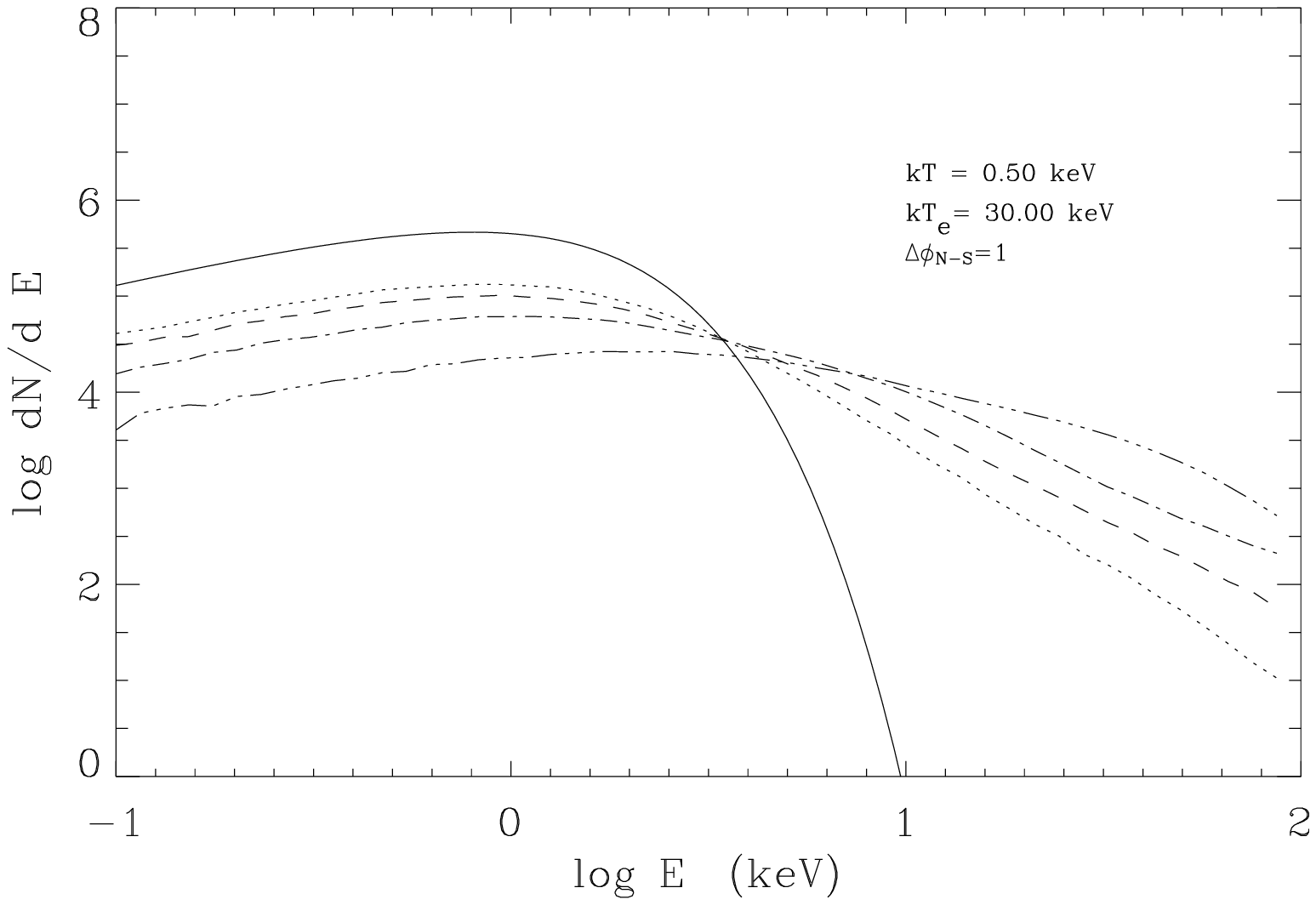}
}
\caption{Top: Computed spectra for $B = 10^{14}~G$, $kT  = 0.5$~keV, $kT_e =
30$~keV, $\Delta \phi = 1$ and different values of $\beta_{bulk}$: $0.3 $ (dotted),
$0.5 $ (short dashed), $0.7 $ (dash-dotted)
and $0.9$ (dash-triple dotted).
The solid line represents the seed
blackbody and counts have been summed over $\Phi_s$.
The two panels
correspond to two different values of the magnetic colatitude:
$\Theta_s=64^\circ$ (left) and $\Theta_s=116^\circ$ (right).
Seed photons are assumed to be  100\% polarized in the ordinary mode.
Bottom: Same, but for seed photons 100\% polarized
in the extraordinary mode.
\label{f2}
}
\end{figure*}

\begin{figure*}
\hbox{
\includegraphics[width=84mm]{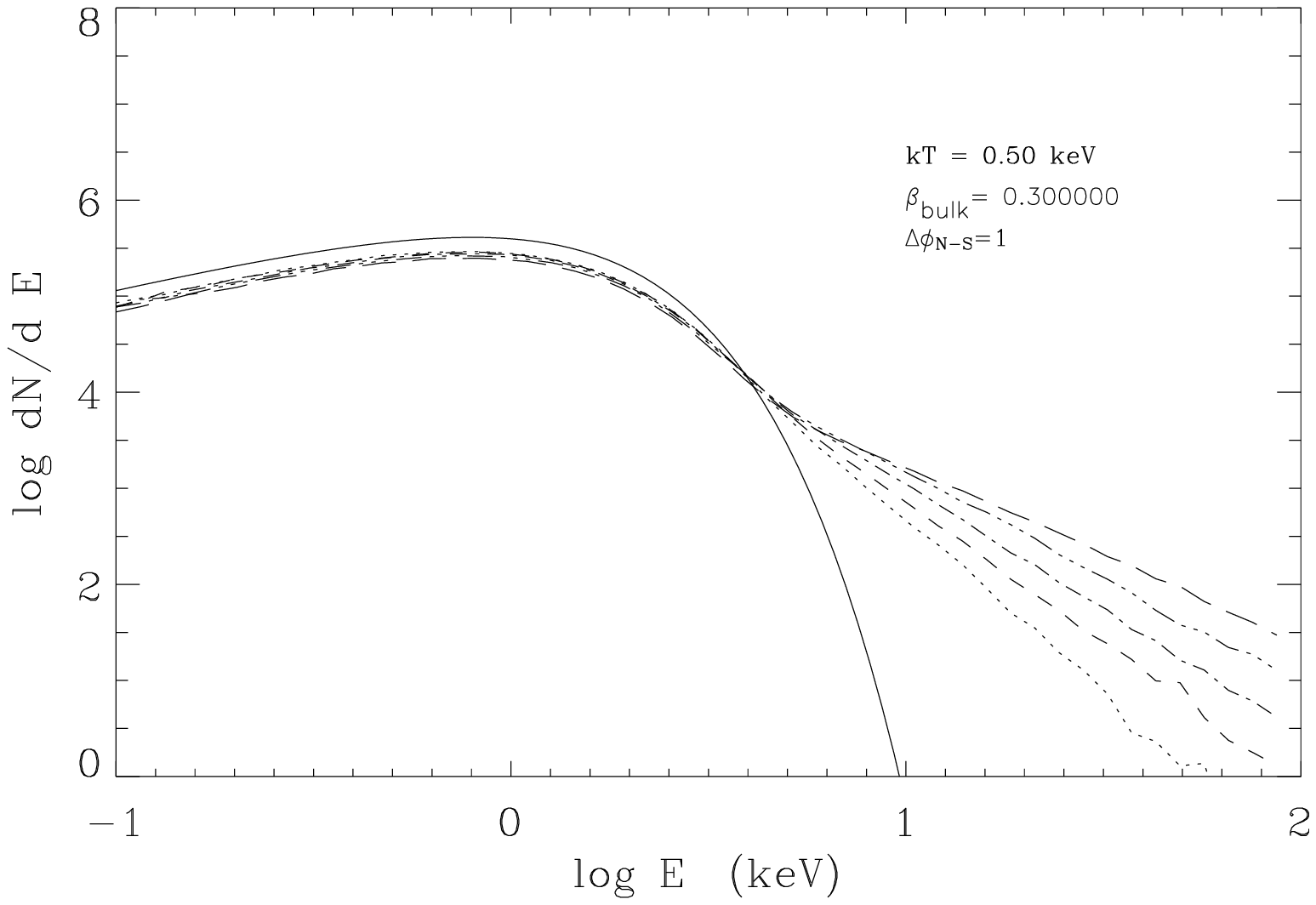}
\includegraphics[width=84mm]{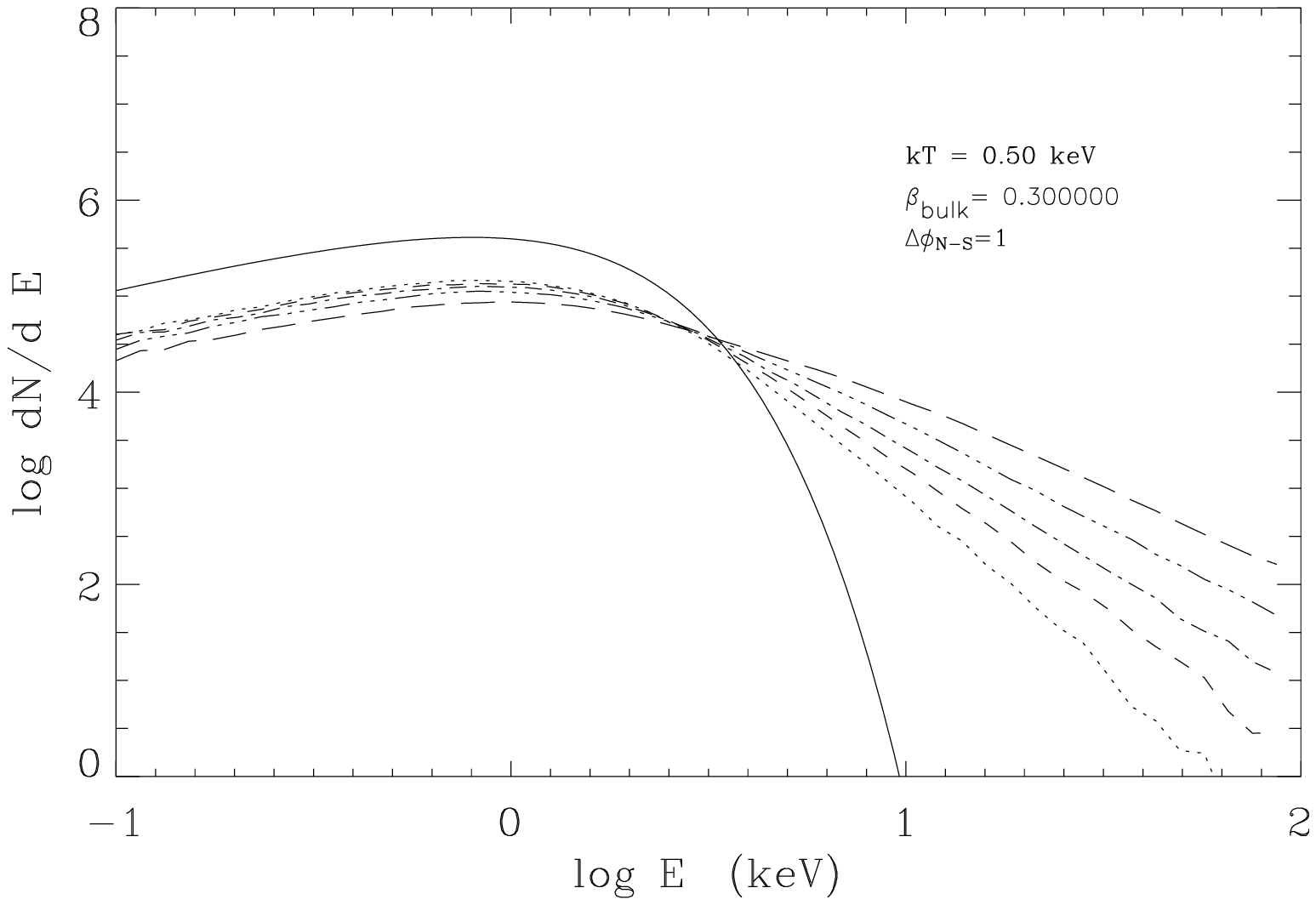}
}
\hbox{
\includegraphics[width=84mm]{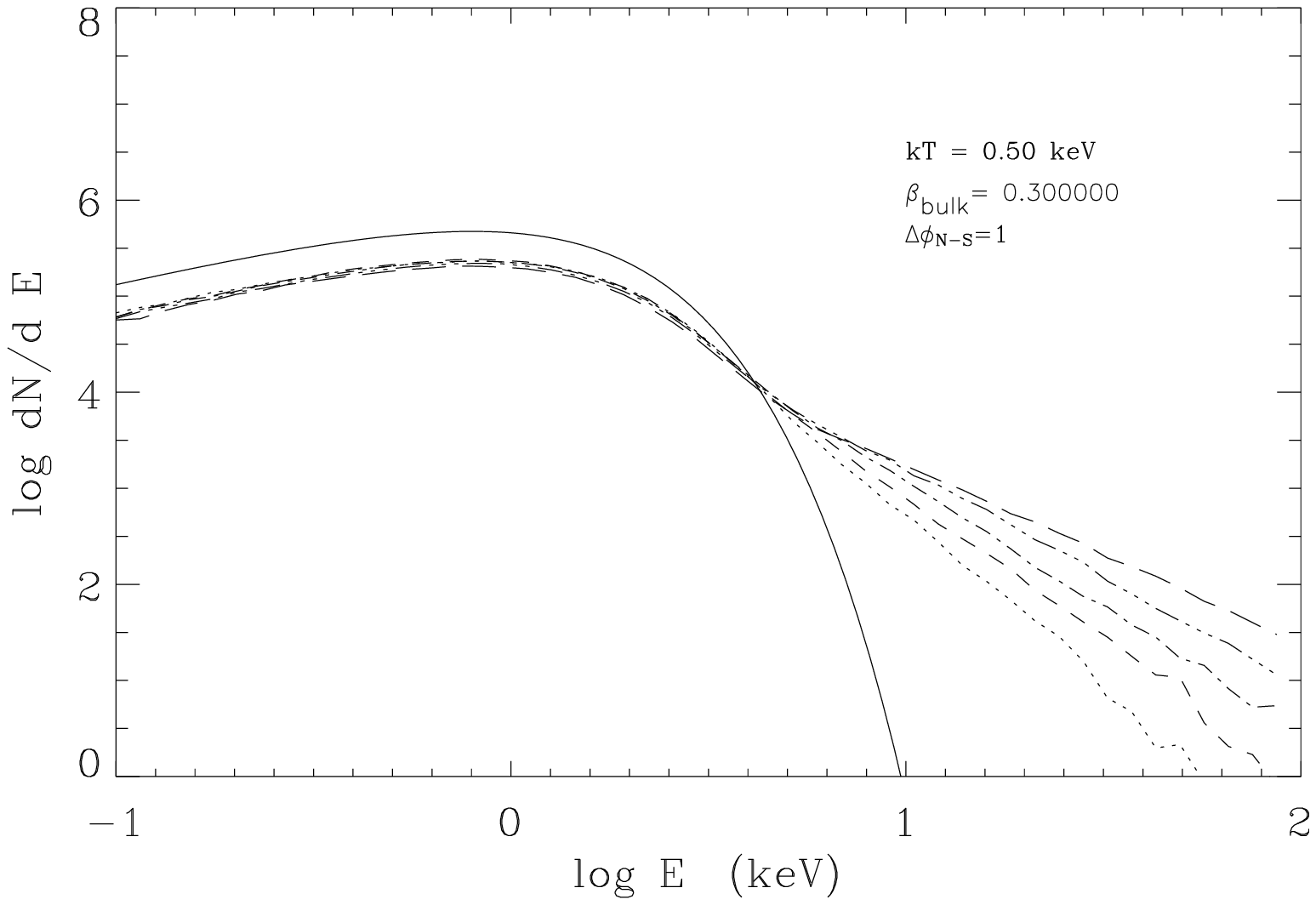}
\includegraphics[width=84mm]{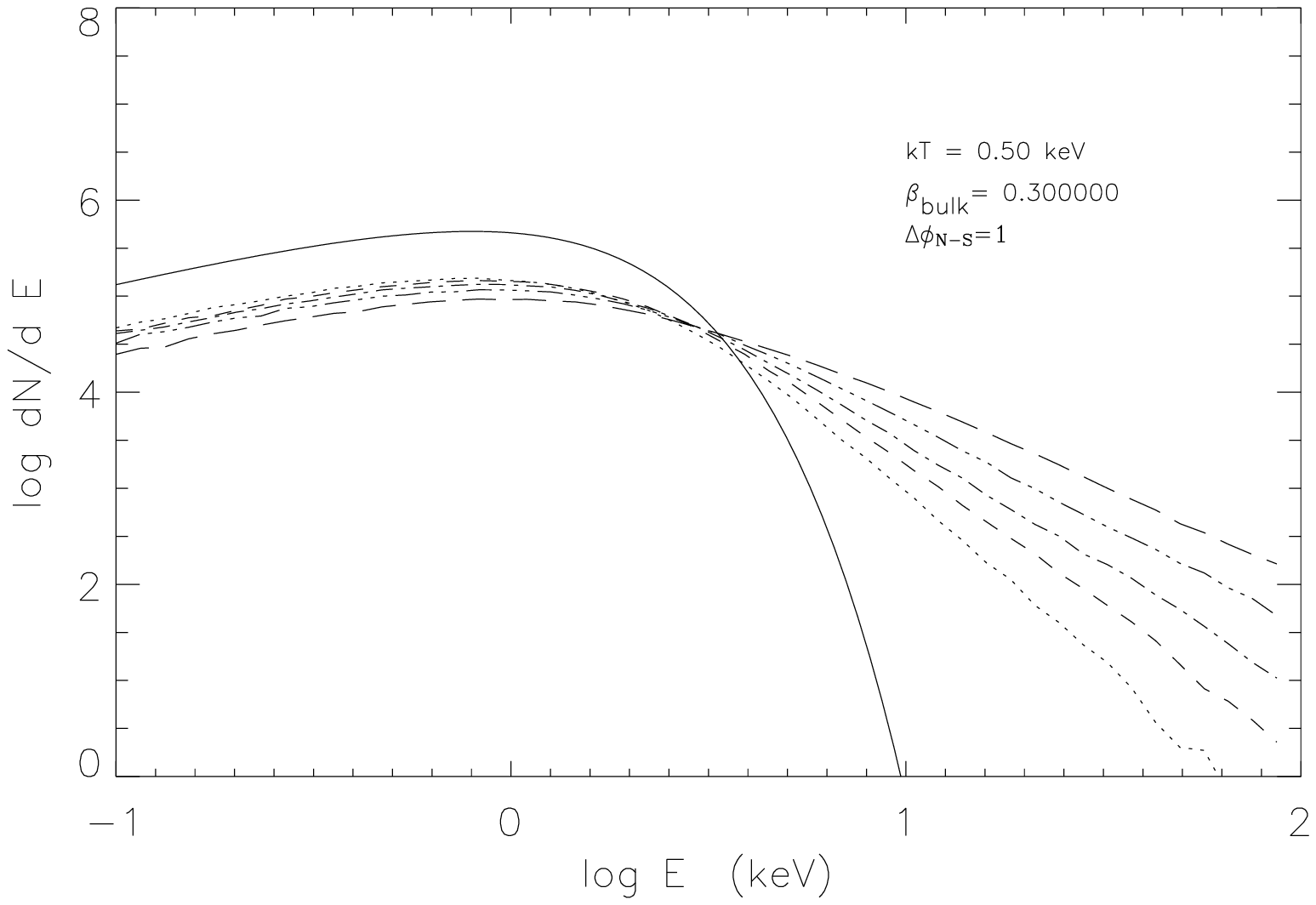}
}

\caption{Top:
Computed spectra for $B = 10^{14}~G$, $kT  = 0.5$~keV, $\beta_{bulk} =
0.3$, $\Delta \phi = 1$ and different values of $kT_e$: 5 keV (dotted),
15 keV (short dashed), 30 ~keV (dash-dotted), 60 ~keV (dash-triple dotted) and 120 keV
(long dashed).
The solid line represents the seed
blackbody and counts have been summed over $\Phi_s$.
The two panels
correspond to two different values of the magnetic colatitude:
$\Theta_s=64^\circ$ (left) and $\Theta_s=116^\circ$ (right).
Seed photons are assumed to be  100\% polarized in the ordinary mode.
Bottom: Same, but for seed photons 100\% polarized
in the extraordinary mode.
\label{f3}
}
\end{figure*}

\begin{figure*}
\hbox{
\includegraphics[width=84mm]{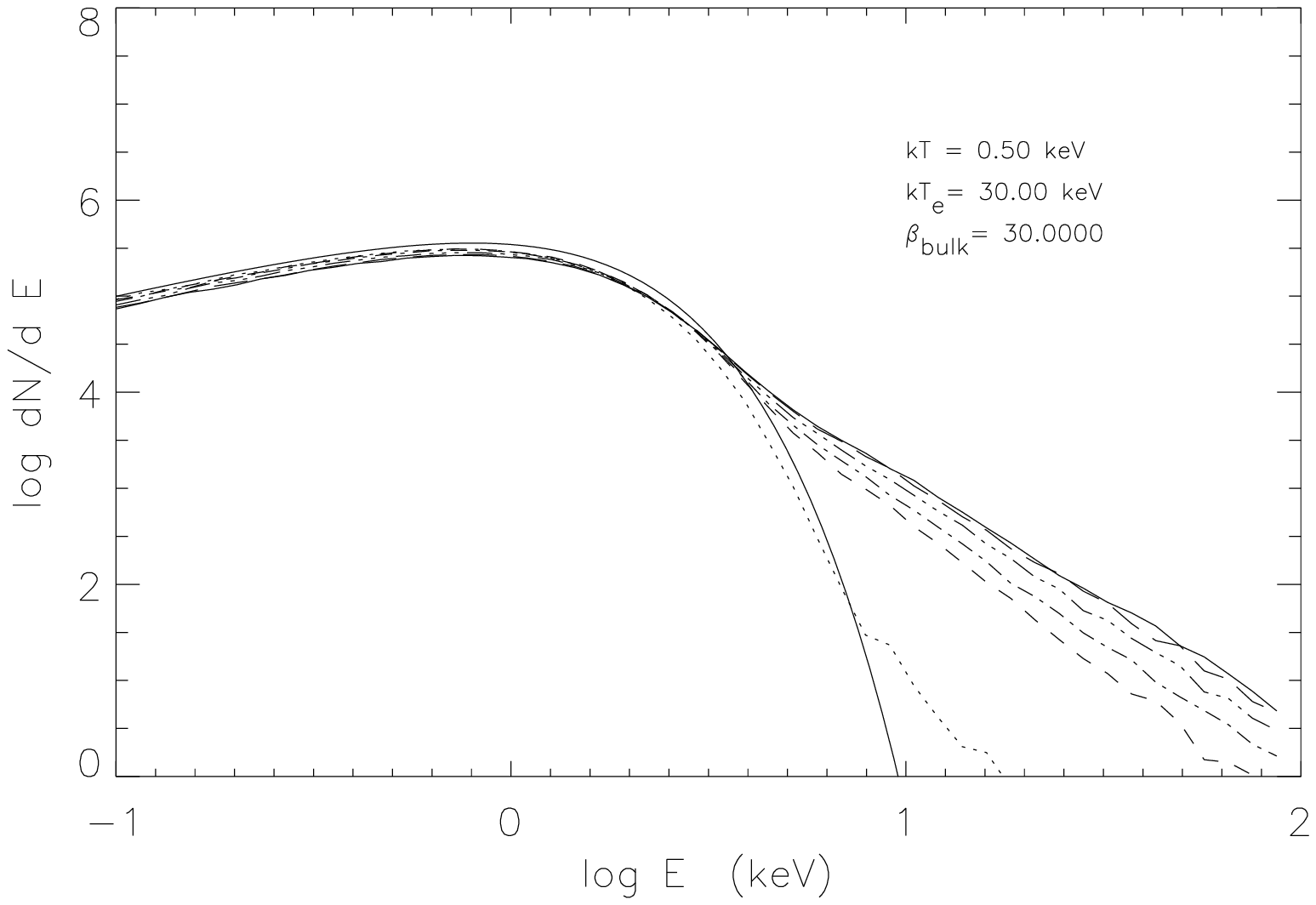}
\includegraphics[width=84mm]{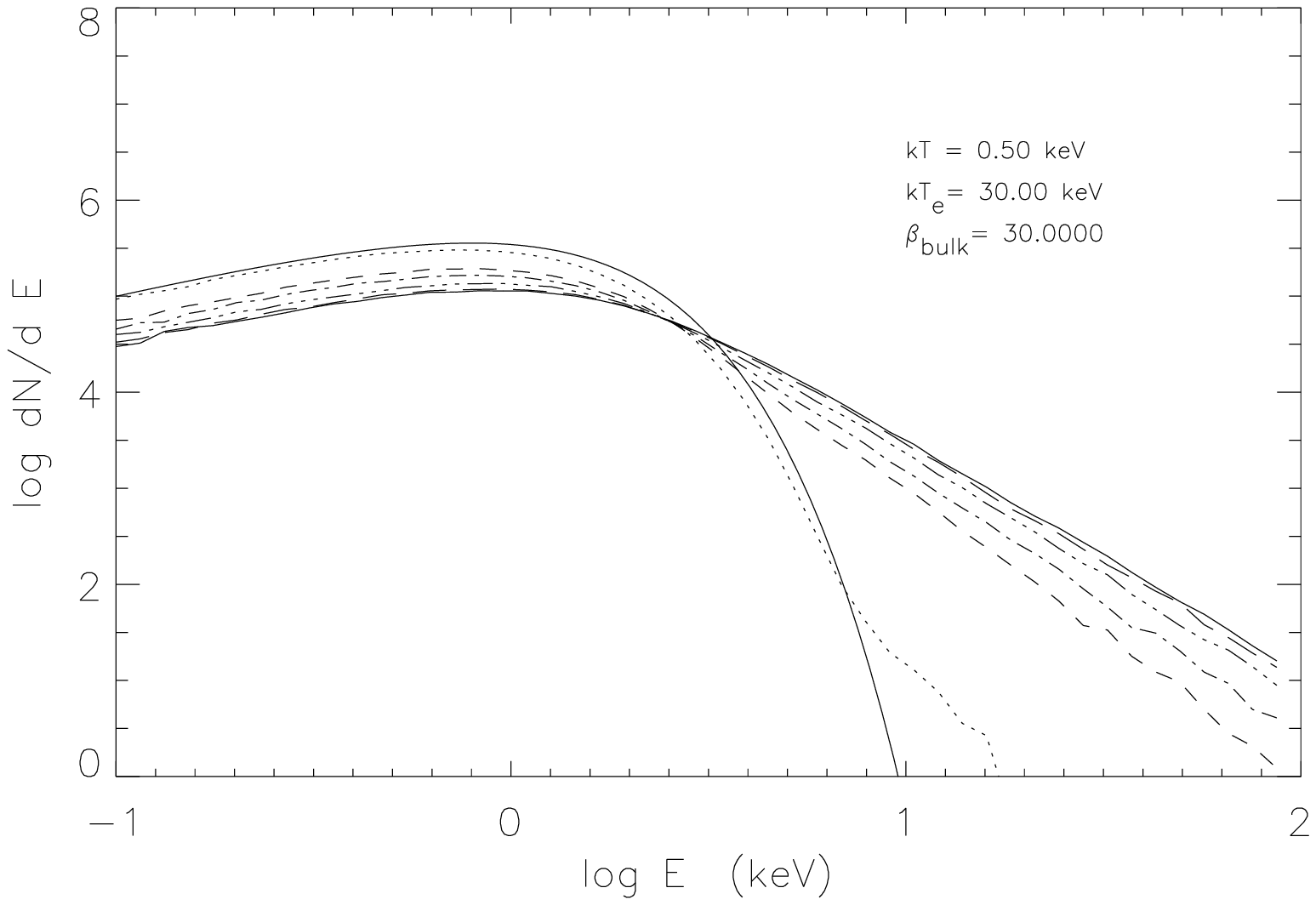}
}
\hbox{
\includegraphics[width=84mm]{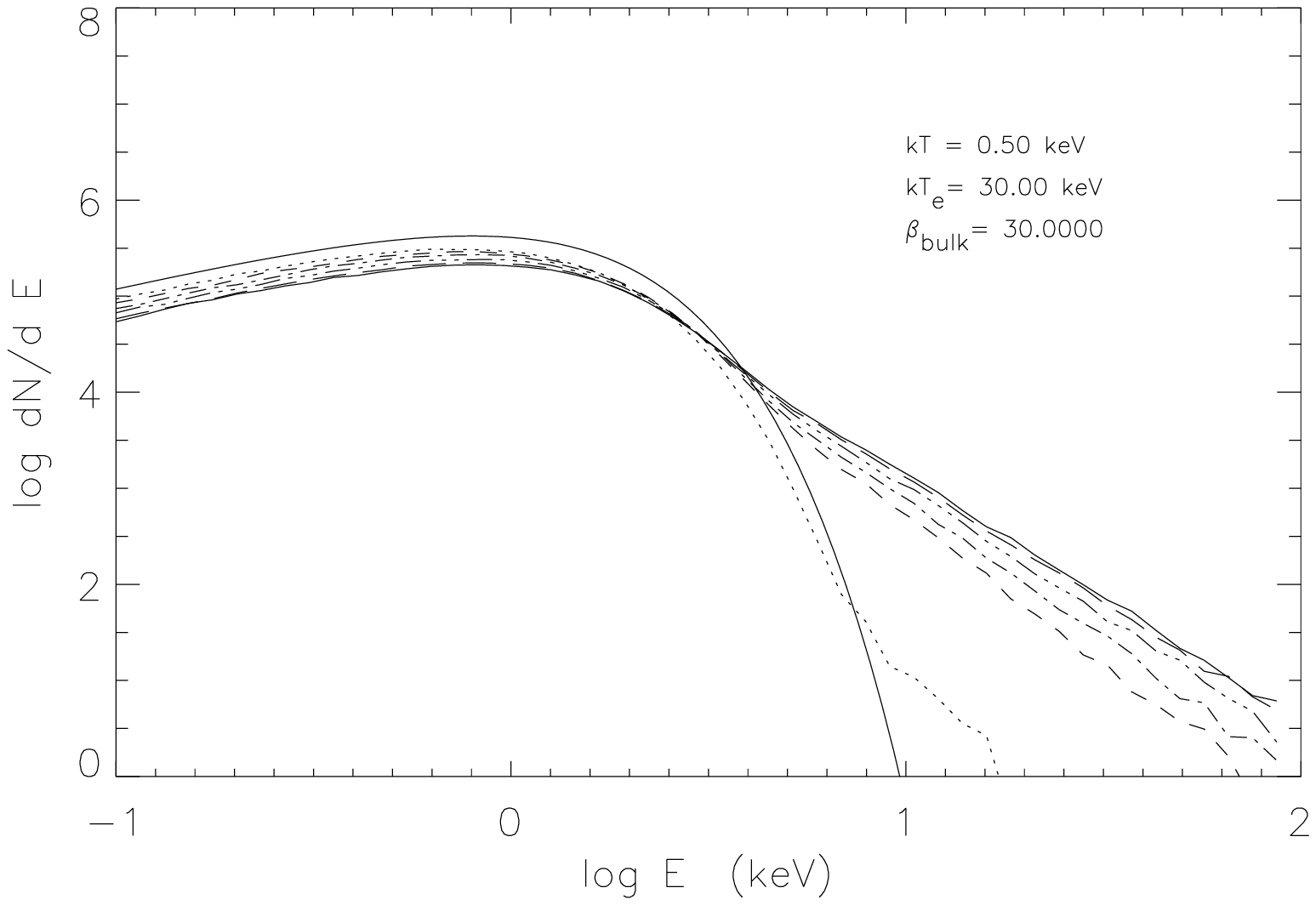}
\includegraphics[width=84mm]{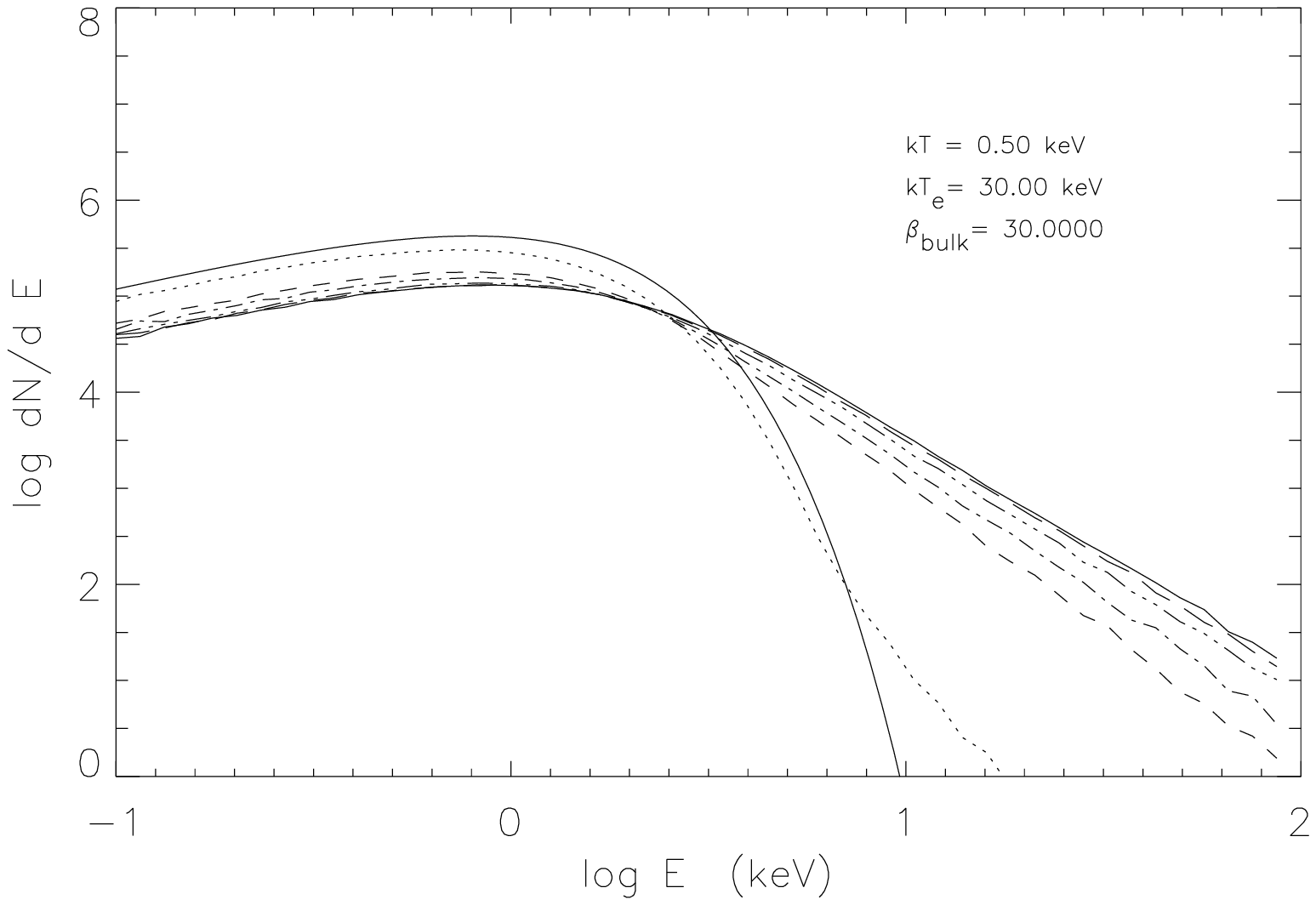}
}

\caption{Top:
Computed spectra for $B = 10^{14}~G$, $kT  = 0.5$~keV, $kT_e =
30$~keV, $\beta_{bulk} = 0.3$ and different values of
$\Delta \phi$: 0.3  (dotted), 0.5  (short dashed), 0.7 (dash-dotted),
 0.9 (dash-triple dotted),  1.1  (long dashed) and
$ \Delta \phi =1.2 $ (solid line, top).
The solid line at the bottom represents the seed
blackbody and counts have been summed over $\Phi_s$.
The two panels
correspond to two different values of the magnetic colatitude:
$\Theta_s=64^\circ$ (left) and $\Theta_s=116^\circ$ (right).
Seed photons are assumed to be  100\% polarized in the ordinary mode.
Bottom: Same, but for seed photons 100\%
polarized
in the extraordinary mode.
\label{f4}
}
\end{figure*}

\begin{figure*}
\includegraphics[width=84mm]{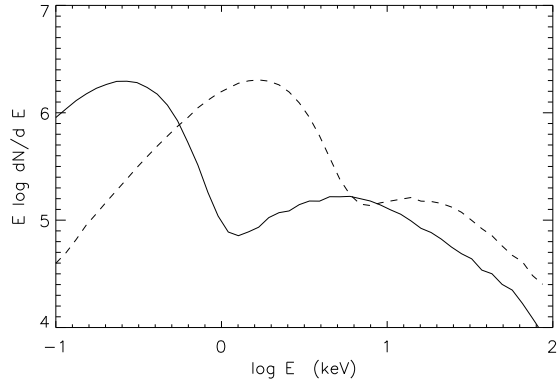}
\caption{Computed spectra for
$B = 10^{14}~G$, $\Delta \phi = 2$; the star is an aligned rotator seen north pole-on. Solid line:
$kT  = 0.1$~keV, $\beta_{bulk} = 0.7$; dashed line:
$kT  = 0.6$~keV, $\beta_{bulk} = 0.6$. In both cases
$kT_e$  is related to $\beta_{bulk}$ through eq.~(\ref{recipe}); seed photons
are unpolarized.
\label{bigobbo}
}
\end{figure*}

Figs.~\ref{f2}, \ref{f3}, \ref{f4} illustrate the effects on the spectral
shape of varying $\beta_{bulk}$,
$kT_e$ and $\Delta \phi$, respectively (here and in the following
we put $\Delta \phi\equiv \Delta \phi_{N-S}$ to simplify the notation). Spectra have been averaged over
$\Phi_s$, and plotted for two values of $\Theta_s$, one
for each hemisphere (left and right panels). As it can be seen, an increase
in each of these parameters (either $\beta_{bulk}$,
$kT_e$ or $\Delta \phi$) always corresponds to an
increase in the comptonization degree of the spectrum. The effect is particularly
notable in the case of $\beta_{bulk}$. If $\beta_{bulk}\ga 0.5$ an observer located in the southern
hemisphere (i.e. with currents flowing towards him)  sees a spectrum
which is no more peaked at $\sim kT$, but peaks instead at about the thermal energy of the scattering
particles. This is because comptonization starts to saturate and photons fills the Wien peak of
the Bose-Einstein distribution. For intermediate values of the parameters,  spectra can
be double humped, with a downturn between the two humps (a clear example of this behaviour is illustrated
in Fig.~\ref{bigobbo}). We note that some of the model spectra presented by
Fernandez \& Thompson (2007) also exhibit a downward break in the tens of
keV range. In particular, when assuming a (non-thermal) top-hat or a broadband
velocity distribution for the magnetospheric charges, they found that multiple
peaks can appear in
the spectrum. The difference is that our model predicts at most two
peaks, and that the energy  of the second one  gives  a direct information
on the energy of the magnetospheric particles. As noticed by \cite{ft07} and
\cite{esp07}, double peaked spectra may play a role in
the interpretation of the broadband X-ray spectrum of SGR 1900+14 and
SGR 1806-20. In particular,
the detection of a spectral break at about a few tens of keV may have
remarkable physical implications and provide important diagnostics for the
physical parameters of the model. A spectral break at $\sim 15$~keV, as
the one possibly detected in the case of SGR 1806-20,
would translate then in a
temperature of $\sim 5$ keV for the magnetospheric electrons
(\citealt{esp07}).

The efficiency of the resonant scattering also increases by
increasing $kT_e$ (Fig.~\ref{f3}), although this effect is less pronounced
than that
observed while increasing the current bulk velocity. This is expected,
because a change in $T_e$ corresponds to a change in the average thermal
velocity for the magnetospheric particles, and not to a boost that
equally affects each single particle. Similarly goes for $\Delta \phi$,
which effect is less pronounced than that of the bulk velocity (see
Fig.~\ref{f4}).
Again, we find that no significant spectral change occurs
exchanging the polarization  of the seed photons from ordinary to
extraordinary.

Although it would be inappropriate to define the RCS spectra as a ``blackbody plus power-law'' (the double-humped
spectra shown in Fig.~\ref{bigobbo} are definitely far away from such a definition), in many cases
the general shape of the continuum is that of a thermal bump and a high-energy tail. In this sense model spectra
are reminiscent of the empirical blackbody plus power-law model often  used to fit (rather successfully)
the magnetars soft X-ray emission. Since, when present, the high-energy tail is indeed power-law-like, it
is of interest to investigate how the spectral index $\Gamma$ (as derived by fitting the high-energy tail
with a power-law)
changes with the parameters. In particular, a hardening of the spectrum is expected for increasing
twist angle (TLK) and this was invoked as a possible mechanism to explain the correlated flux-hardening
variations in some sources \cite[e.g.][]{sandro05a, rea05}. This is confirmed by our calculations
\cite[see also][]{ft07}, as shown in Fig. \ref{fgamma}. The photon index monotonically decreases
with $\Delta\phi$, going, in the present case, from $\sim 3$ to $\sim 2.4$ by changing the twist
angle by $\sim 1$ rad. The behaviour is quite similar at both the field
strengths we considered, although spectra for $B=10^{15}$ G are fractionally harder. The model shown
here has $kT_e = 30$ keV, $\beta_{bulk} = 0.3$, a uniform surface temperature $kT = 0.5$ keV and spectra
have been obtained summing over all the sky patches.

To illustrate the effects of a non-homogeneous surface temperature distribution, we discuss the case in which
photons are emitted by a single patch. The subdivision of the star surface and of the sky
is the same as that adopted before, and also the energy range and bin width. In the present run
the seed radiation is taken to be isotropic ($b = 1$) and unpolarized, i.e. an equal number of
ordinary or extraordinary photons are emitted, and, again, the spin and magnetic axes coincide.
We selected an  emitting patch located just above the equator
(centred at $\Theta=77.5^\circ,\ \Phi=45^\circ$)
with a surface temperature of $kT=0.5$ keV.
The magnetospheric parameters are $\beta_{bulk}=0.3$, $kT_e=10$ keV and $\Delta\phi=1.3$. Figure
\ref{onepatch} shows the emerging spectrum, as viewed by an observer whose LOS
makes an angle $\Theta_s=90^\circ$ with the spin axis (i.e. the star is
seen equator-on) for different
values of the observing longitude, $\Phi_s=20^\circ,\, 140^\circ,\,
220^\circ$. These three values
correspond to having the emitting patch in full view (seen nearly face on), partially in view and screened
by the star. The effects of the different viewing angle on the spectrum are dramatic. When the
emitting patch is in full view both the primary, soft photons and those which undergo repeated
resonant scattering reach the observer and the spectrum is qualitatively similar to those presented
earlier on, with a thermal component and an extended power-law-like tail. On the other
hand, if the emitting region is not directly visible, no contribution from the primary blackbody photons
is present. The spectrum, which is made up only by those photons which after scattering propagate
``backwards'', is depressed and has a much more distinct non-thermal shape.

\subsection{Polarization of the emitted radiation}
\label{res_pol}

Radiation emerging from strongly magnetized neutron stars is expected to be highly polarized, due to the strong
dependence of radiation transport on the photon propagation mode. Polarization studies have already started at
low energies (IR), and future X- and $\gamma$-ray polarimetry with high sensitivity instruments, such as the
planned photoelectric polarimeter to be flown on the ESA mission {\em XEUS}, are expected to extend them over a
broader spectral band. The development of detailed theoretical predictions is therefore fundamental: polarimetry
will bring into view  a new and unique dimension of the problem, through the knowledge of polarization degree
and swing angle.

In our scenario, the degree of polarization in the soft X-ray
radiation emitted by magnetars results from a combination of several
effects. Seed thermal photons,  originating from the crust or atmosphere of
the star, do posses an intrinsic polarization \citep[e.g.][]{sil00,vanlai06}.
The fraction of polarization, which is
determined by the competition between plasma and vacuum properties,
depends on the energy band, and on the details of the density
and temperature gradient in the emitting region. Seed photons then
propagate in the magnetosphere, where multiple resonant scatterings
further influence the polarization degree. By using our Monte Carlo simulation,
we are in the position to investigate the latter effect, i.e. to estimate the
degree of polarization which is expected to arise because of
magnetospheric effects only and to investigate its dependence on the model
parameters.

In Figs.~\ref{polla1} and \ref{polla2} we show, as a function of various
parameters, the degree of polarization of the emerging radiation,
defined as $\vert N_{extr} - N_{ord}\vert/(N_{extr} + N_{ord})$ where
$N_{extr}$ and $N_{ord}$ are, respectively, the number of ordinary and
extraordinary photons collected at infinity. The polarization degree has
been averaged over frequency, over the whole emitting surface and over the
sky at infinity. As it can be seen, the efficiency at
which completely polarized surface radiation is de-polarized increases by increasing the strength
of magnetospheric upscattering, i.e. by increasing one of the three parameters
$\beta_{bulk}$, $kT_e$ or $\Delta \phi$. This effect
is stronger for ordinary seed photons, for which the probability of undergoing mode
switching in the scattering process is higher (see, e.g.,  eqs. [\ref{totcrosslab}]) and
 for photons emitted close to the south pole (see
Fig.~\ref{polla3}, the latter result reflects our choice for the direction
of the current flow, as discussed earlier). On the other hand, would the
surface radiation be completely unpolarized, we can see that, while
passing through the magnetosphere,  it can
acquire only a relatively small degree of linear polarization: typically
10--20\%, up to 30\% for very extreme values of the current  bulk velocity.
This means that, would future observations of X-ray polarization result in
measurements larger than 10-30\%, the excess has to be attributed to an
intrinsic property of the surface radiation.

We have also explored how the polarization degree depends on the photon
energy and
a representative case is shown in Fig. \ref{fpolfreq}. The two panels
refer to a run with the same set of model parameters ($\beta_{bulk}=0.3$,
$kT_e= 10\, {\rm keV}$, $kT=0.5 \, {\rm keV}$, $\Delta\phi=1.3$) but performed assuming
that seed photons are completely polarized either in the extraordinary (left
panel) or ordinary mode (right panel). The
polarization degree has been computed as above, but now different viewing
directions are retained (i.e. only sum over $\Phi_s$ has been performed).
Emission is again from the entire surface (at constant $T$) and the star is an aligned rotator.
As expected, for 100\% polarized seed photons the polarization degree decreases with
increasing energy,  since harder photons undergo more scatterings. Low energy photons tend to keep
their original polarization
state, although there is a dependence on the viewing angle. Not surprisingly, even at low energies, the
polarization degree is higher when the LOS is close to the north pole (dash-triple dotted lines in Fig.
\ref{fpolfreq}) and drops for increasing viewing angle. It is interesting to note that the largest
de-polarization (at low energies) does not occur close to $180^\circ$ but when viewing the star southern hemisphere
at an intermediate angle because of the low particle density near the poles.

\subsection{Viewing angle effects}
\label{view}

Spectra presented in \S\ref{spect} have been computed
accounting for different viewing angles only in the case in which the
star is an aligned rotator, i.e assuming that the spin and magnetic axes
coincide. Under this hypothesis, the viewing geometry is described by a single angle
which is just the colatitude $\Theta_s$ of the centre of the sky patch where photons
are collected. Since the magnetic field and the current distribution are axially symmetric,
the contributions from {\em all} the sky patches located at the same value of $\Theta_s$ (and
different $\Phi_s$) may be summed together if surface emission is itself axisymmetric, as in the uniform 
temperature case discussed at the beginning of this section.

In order to treat the more general case in which the spin and magnetic axes are not aligned,
we introduce two angles, $\chi$ and $\xi$, which give, respectively, the
inclination of the LOS and of the dipole axis with respect to
the star spin axis. This also allows us to take into account for the star rotation and
hence derive pulse shapes and phase-resolved spectroscopy. Because of the lack of north-south
symmetry, it is $0\leq\chi\leq\pi$, while $\xi$ spans the interval $[0,\,\pi/2]$. By introducing
the rotational phase $\alpha\ (0 \leq \alpha \leq 2 \pi)$, the co-ordinates of the point
which represents the intersection of the LOS with the sky for each value of $\alpha$ are

\begin{eqnarray}
\label{angoli}
\cos \overline \Theta_s &=& \cos \chi
\cos \xi + \sin \chi \sin \xi \cos  \alpha \\
\nonumber
\cos\overline
\Phi_s &= &\frac{\cos  \chi - \cos \overline \Theta_s \cos \xi}
{\sin \overline \Theta_s \sin \xi }\, .
\end{eqnarray}

At constant $\chi$ and $\xi$, eqs. (\ref{angoli}) trace a circle on the sphere which represents the sky.
As a result of each Monte Carlo run, the spectrum in counts has been recorded for each pair
of values $\Theta_{s,i},\,\Phi_{s,j}$ which correspond to the centres of the sky patches,
$N(\Theta_{s,i},\Phi_{s,j}, E_k)$. In order to compute the spectrum at a discrete set of phases $\alpha_l$, we
perform a double interpolation of this array over the angular variables, to obtain the number
of counts in correspondence to the pair of angles $\Theta_s(\alpha_l),\, \Phi_s(\alpha_l)$ given by
eq.~(\ref{angoli}), i.e. $N_{ph}(\alpha_{l}, E_k)$. Finally, integration of $N_{ph}$ over $E$ or $\alpha$
gives the lightcurve in a given energy band, or the phase-averaged spectrum, respectively.
An illustration of the effects of a different viewing geometry is shown in
in Fig.~\ref{variachi}, where spectra correspond to increasing values of
$\chi$.

A systematic investigation of the properties of the pulse shape while varying the
model parameters is beyond the purpose of the present paper, and it will be
presented elsewhere (Albano et al. in prep.). Here we just show in Fig. \ref{figlc} two
examples, both relative to a star seen equator-on ($\chi=90^\circ$), but for two different
inclinations of the magnetic axis ($\xi=10^\circ$ and $\xi=50^\circ$). In the first case the
pulse profiles in the soft (0.5--2 keV) and hard (2--6 keV) band are shifted in phase by $\sim 180^\circ$.
By increasing $\xi$ the pulsed fraction and the pulse shape sensibly change
with the energy band. The pulsed fraction increases with the energy and, at the same time, the
double peaked structure present in the low energy band disappears at higher energy where the
lightcurve is sinusoidal.

\section{XSPEC implementation and applications}
\label{arch}

One of the goals of the present investigation is to apply the resonant compton scattering
model discussed in the previous sections to magnetar spectral fitting, by implementing it into
the standard package for X-ray spectral data analysis XSPEC. Clearly, our Monte Carlo
spectra can be loaded in XSPEC only in tabular form, using the {\tt atable} option. This implies
that a model archive has to be generated beforehand, for a reasonably wide range of the model parameters.
Although a production run takes (under typical conditions) about 20 m, building a large model archive
necessary demands for a compromise between generality, accuracy and feasibility. As we discussed already
(see \S\ref{spect}), the model has four parameters: $\beta_{bulk}$, $kT_e$, $\Delta\phi$ and $kT$, assuming
that the surface is at constant temperature. If a model is computed for, say, ten values of each parameter,
this would result in a total of $10^4$ runs requiring about $2\times 10^5\ {\rm m} \sim 140\ {\rm d}$ of CPU
time. Even splitting the computation over a few machines, the time needed ($\sim$ month) is barely
acceptable. Moreover, we are aware that the adopted description of the charge velocity distribution,
which involves two out of four model parameters, is far from
being consistent. For these reasons, we decided
to simplify our treatment by imposing that the electron bulk kinetic and thermal energies are related.
The mean thermal energy for a 1D relativistic Maxwellian distribution can not be expressed in closed form.
However, to an excellent accuracy, it is

\begin{equation}
\label{meantherm}
\langle\gamma-1\rangle\simeq\frac{\Theta_e}{2^{1/(1+\Theta_e)}}\,.
\end{equation}
We then derive the value of the electron temperature by assuming equipartition between
thermal and bulk kinetic energy, i.e. by solving for $\Theta_e=kT_e/m_ec^2$ the equation
\begin{equation}
\label{recipe}
\gamma_{bulk}-1=\frac{\Theta_e}{2^{1/(1+\Theta_e)}}\,.
\end{equation}
In order to avoid that for the higher values of $\beta_{bulk}$ we consider (see below) the assumption
of conservative scattering in the ERF is invalidated, the solution of eq. (\ref{recipe}) is actually halved.

The grid of models has been generated for $0 \leq \Delta \phi \leq 2$ (step 0.1),
$0.1 \leq \beta_{bulk} \leq 0.9$ (step 0.1) and eight values of $kT$ (in keV): 0.1, 0.13,
0.16, 0.2, 0.25, 0.40, 0.63, 1, under the
assumptions that the surface has a constant temperature, emits
isotropically and the surface radiation is unpolarized. The number of
divisions on the star surface and on the sky, the
energy range and bins are taken as in \S~\ref{spect}, but now we evolve
$N_{patch}=225,000$ photons per patch, therefore each model corresponds to
$N_{tot}= 7,200,000$ photons. Again, the magnetic field is fixed at
$B=10^{14}$~G (further archives corresponding to different values of $B$
can be easily generated).
The computed spectra have then been averaged over the  whole sky at infinity,
smoothed and re-interpolated (using a logarithmic interpolation) over a grid of 300
equally spaced energies in the range
0.1--15 keV and on a logarithmic grid of 100 equally spaced temperatures in
the range $-1 \leq \log kT \leq 0$. The latter step is necessary because interpolation
on the logarithm of the spectrum with respect to parameters is not possible within XSPEC
for tabular models. After some experimenting, we found that in order
to have enough accuracy when interpolating the spectrum a fine grid in $kT$ is necessary.
The final XSPEC {\tt atable} spectral model (22~MB in size,
named {\tt ntznoang.mod}) has been created by
using the routine {\tt wftbmd}, available on-line.\footnote{see
http://heasarc.gsfc.nasa.gov/docs/heasarc/ofwg/docs/general/
\\ modelfiles\_memo/modelfiles\_memo.html.}

The {\tt ntznoang} model has four free parameters ($\beta_{bulk}$, $\Delta \phi$, $\log kT$
plus a normalization constant), which can be  simultaneously varied
during the spectral fitting following the standard $\chi^2$ minimization
technique. It is important to note that this model has the same number of
free parameters than the canonical blackbody plus power-law empirical
model or the RCS model recently discussed in \cite{nanda2}, and hence has
the same statistical significance.

Following essentially the same procedure outlined above and making use of the
same archive, we have also built a XSPEC model in which the dependence on the
two geometrical angles, $\chi$ and $\xi$, is explicitly accounted
for, as discussed in \S\ref{view}. Phase-averaged spectra have been computed
on a $7\times7$ equally-spaced grid of $\chi$ and $\xi$ values. The two angles
are in the in the ranges $ 0\leq \chi \leq 180^\circ$ and $0 \leq \xi\leq 90^\circ$,
respectively. At variance with the angle-averaged case considered previously, the
grids in the other parameters (except $kT$) are coarser: $0 \leq \Delta \phi \leq 1.8$ (10 values, step 0.2),
and  $0.1 \leq \beta_{bulk} \leq 0.9$ (5 values, step 0.2). Maintaining the same parameter grids used
to build the {\tt ntznoang} model would, in fact, result in too a large file to be read into
XSPEC. The final {\tt atable} spectral model, {\tt ntzang.mod},
is $\sim300$ MB in size, and has six free parameters ($\beta_{bulk}$,
$\Delta \phi
$, $\log kT$, $\chi$, $\xi$
plus a normalization constant). Despite the larger number of free parameters, the ``angular'' model
can be used to infer information about the viewing geometry, eventually combining
information that can be obtained  by fitting simultaneously
phase-resolved spectra, or independently from the
study of the pulse profile.

A systematic application of both models to magnetars spectra is in
progress, and will be reported elsewhere (Israel et al. and Rea et al. in preparation).
Here we present only an example which is illustrative of how the two {\tt atable}
spectral models behave when applied to X-ray data. Fig.~\ref{noang} (left panel) shows the fit of the
0.1--10~keV {\it XMM-Newton} EPIC-pn spectrum of the transient AXP \wes\ taken on February 17
2007, i.e. about five months after a burst and a glitch were detected from this source
\cite[][see also \citealt{muno07, giallo07}]{cxo, giallo06}.
All details about the observation will be reported in Israel
et al.~(in preparation). The spectrum has been modeled with the angle-integrated
{\tt ntznoang} model, modified by interstellar absorption ({\tt phabs}
model in XSPEC). Data and best fitting model are shown in Fig. \ref{noang} and
the best fit parameters are listed in Table~\ref{tab1}.
As expected, and as it has been also found in other applications of the
RCS model \citep{nanda2}, the inferred value of the column
density is
smaller than that implied by a blackbody plus power-law fit. This is because
the empirical blackbody plus power-law modelling is
known to overestimate the soft X-ray emission and, in turn, the
value of the interstellar absorption.

Since the fit is already very good ($\chi^2_{\nu} = 0.81$), there is no
statistical need to
introduce two further parameters. However, we also tried to fit the same
observation with an absorbed {\tt ntzang} model, with the only goal to
check and test the correctness of its  XSPEC  implementation; results are
shown in Fig~\ref{noang} (right panel) and reported in Table~\ref{tab1}. As expected,
the values of the angles are unconstrained, and the remaining parameters
are
in agreement with those found with the first model. Again, is not our main
scope to provide the physical values of the angles here: instead  we
stress that this figure is presented purely as an illustration.
Nevertheless, the successful spectral fit with the {\tt ntznoang}
model clearly demonstrates
that the model can catch the main
features of the magnetar emission and reproduce them quantitatively.

\section{Discussion and conclusions}
\label{disc}

In this paper we have investigated how the thermal spectrum emitted by the
star surface gets distorted by repeated resonant scatterings onto mildly
relativistic magnetospheric electrons using a Monte Carlo technique. The goal
of this study has been twofold. Our first motivation has been to create a model
archive which  could be
implemented as a tabulated model in XSPEC and directly
applied to fit the spectra of magnetar candidates. The model is available
in two versions, with or without the explicit dependence on the two angles
which give the inclination of the line-of-sight and the magnetic axis wrt
the star spin. A systematic application to different sources is under way
and here a (preliminary) fit to the {\em XMM-Newton}
spectrum of \wes has been presented, mainly for illustrative purposes.

In building our Monte Carlo code we have followed an approach similar
to that discussed in \cite{ft07}. However, the two codes differ in many
respects. A major difference is in the adopted description of the velocity
distribution of the scattering particles. We have
explicitly accounted for the collective (bulk) electron motion associated
to the charge flow in the magnetosphere, superimposed to which we assume a 1D
relativistic Maxwellian distribution which simulates the particle velocity
spread.  We also allow for a completely general description of the star surface thermal
map and this makes it possible to assess the effects of a (spatially)
localized emission (e.g. by a hot spot). Moreover, in our treatment seed
photons are not taken to move only in the radial direction
but are drawn from a prescribed angular distribution which can account for magnetic
beaming effects.

As the present application to \wes\ shows
\cite[\S\ref{arch}; see also][]{lg06, nanda1, nanda2},
spectral models based on resonant cyclotron up-scattering of thermal
photons in the magnetosphere of magnetars prove quite successful in
interpreting quantitatively the soft ($\sim 1$--10 keV)
emission from AXPs and SGRs. Albeit the numerical computation presented here includes several
important details about the microphysics and the magnetospheric properties
and geometry, it relies on some simplifying assumptions which reflect our
poor knowledge on some key issues of magnetar physics.

A prominent one is the nature of
the plasma which fills the magnetosphere. Most investigations on RCS, including our, restricted
to unidirectional flows, i.e. assumed that scattering occur onto electrons \cite[a simple bi-directional
flow was considered by][]{ft07}. As discussed by \cite{beth07}, in a twisted magnetosphere charges,
accelerated by the self-induction electric field, may produce $e^\pm$. 
Pairs definitely contribute to
the scattering depth
\footnote{As discussed by \cite{ml07}, for an iron crust  
and magnetic fields as high as $\sim 10^{15}$G, 
vacuum gaps may be formed above the 
polar regions of SGRs/AXPs, with subsequent pair creation. The
pair-dominated region, however, is very thin
and located just above the star surface. This
implies that scattering is resonant for  photon energies in the tens
of MeV range. Since thermal emission from the star surface does not supply 
such high-energy photons, pair cascades produced by 
the gap breakdown are not going to affect our results.}. The 
final spectral shape depends on which 
species populate the corona
and on their spatial and velocity distribution. Our choice of modelling the $e^-$ current
in terms of a bulk motion plus a velocity spread seems to be at least in qualitative agreement with
the analysis presented by \cite{beth07}. We point out, however, that the assumption of
a 1D thermal distribution for the particle velocity in the local rest frame is somehow
arbitrary and no attempt has been made here to assess the effects of other possible
(local) distributions. This has been done, in a few representative cases, by \cite{ft07}, who
did not include, however, the charge bulk motion.
By comparing our results with their, one may conclude that, while
the general effects induced by magnetospheric RCS on primary thermal photons (i.e. the
formation of a ``thermal-plus-power-law'' spectrum) are not much sensitive
to  the assumed particle velocity distribution, the details of the  spectral shape do.

A further caveat concerns the star temperature distribution and the primary spectrum.
Our model archive has been generated assuming that the star radiates a blackbody from a uniformly
heated surface. At present it is unclear if magnetars do possess an atmosphere. A possibility
is that highly energetic electrons hitting the surface knock out protons which then sublimate
giving rise to a ``current induced'' atmosphere \citep{beth07}. Departures from a blackbody
primary spectrum due to reprocessing in a strongly magnetized atmosphere are, however, not expected to be
dramatic \cite[see e.g.][]{sil01,holai01,laiho03}. On the other hand, the issue of the surface thermal map
appears more serious since even passively cooling isolated neutron stars are known to have a non-uniform surface
temperature \cite[see e.g.][]{dany95, silrob06}. In the case of a magnetar, returning currents impacting on the
star surface produce localized heating (TLK). Moreover, starquakes, possibly triggered by the strain accumulated
during the growth of the twist and connected to the glitching activity discovered in AXPs \cite[see e.g.][]{simo03},
can further contribute to the injection of heat into limited portions of the crust. Transient AXPs might be
powered in a similar way by the sudden release of energy into a localized area of the star surface, as observations
of the TAXP XTE J1810-197 seem to indicate \citep{eric07}.

Although the twisted dipole model used here has the advantage of simplicity while catching the essential
physical features, most probably it gives only an idealized representation of the magnetic field outside a magnetar.
The twist may be confined at high magnetic latitudes (TLK), or, if global, it might involve magnetic
configurations more complex than a dipole. Possible evidence for a twist which involves in the first place the field
lines closer to the magnetic poles have been discussed by \cite{woods07} in connection with the period derivative
evolution and its correlation with spectral hardness in SGR 1806-20 before and after the giant flare of December 27 2004.

Both \cite{lg06} and \cite{ft07} assumed that scattering is conservative in the electron rest frame. As discussed
in \S\ref{cross-sections} this choice is quite adequate if spectral modeling is restricted to the soft X-ray
range and has been retained in the present work. However,
the X-ray spectra of magnetar candidates  are nowadays known to
exhibit also a high energy ($\sim 20$--200 keV) component, which is completely non-thermal
and is responsible for about half of the bolometric flux.
Although different scenarios for the origin of the high energy emission
from magnetars have been put forward, not necessary involving RCS (see
\S\ref{intro}), an intriguing possibility is that also the hard tail
arises because of resonant upscattering in the magnetosphere
\citep{baring}. Given the much higher photon energies (in the 100 keV
range) this necessary requires the presence of highly relativistic
electrons (pairs), and, consequently, any attempt to model RCS under those
conditions demands for a fully relativistic, QED treatment of the
scattering cross sections. Although we
presented here spectra extending up to 100~keV, they must be
considered as trustworthy only until $\hbar\omega \ll m_ec^2/\gamma$, i.e. up to a few tens of keV.
Above these energies electron recoil starts to become important and the spectrum is expected to
break. The precise localization of the break would come only from
a consistent treatment, and is particularly important to explain the COMPTEL upper limits observed in
some magnetar sources \citep{kuiper, nanda07}. Moreover, if hard tails are due to a secondary
population of ultra-relativistic electrons confined close to the stellar
surface \cite[as proposed by][]{baring}, resonant scattering
would occur at much higher  values of the magnetic field, $B > B_{QED}$,
which makes the need of a completely QED treatment of the cross section
even more necessary. Same holds for computations aimed at assessing the
role of ions in shaping the spectra. As previously discussed {see \S\ref{currents}}, positively
charged ions are expected to populate the
twisted magnetosphere, but whether these particles can effectively shape
the X-ray spectra is mainly related to the role of those ions located
close to the star surface. The inclusion of this effect, however, requires the
knowledge of the full QED resonant cross section for protons/ions which at
present has not been investigated in detail.

Future work needs to address this issue, among others. Clearly, in order to
to include the relativistic treatment of the scattering process in the electron rest frame,
having a tested, reliable Monte Carlo code which can be easily generalized is of fundamental
importance and this has been our second  motivation in undertaking
this study. In order to extend our
computation of resonant electron cyclotron scattering to the relativistic regime,
we are already completing a detailed investigation of the QED resonant cross
section.  This will be then implemented in our
Monte Carlo code and results will be presented
in  forthcoming papers (Nobili, Turolla \& Zane in preparation).

\section*{Acknowledgments}

\begin{figure*}
\includegraphics[width=84mm]{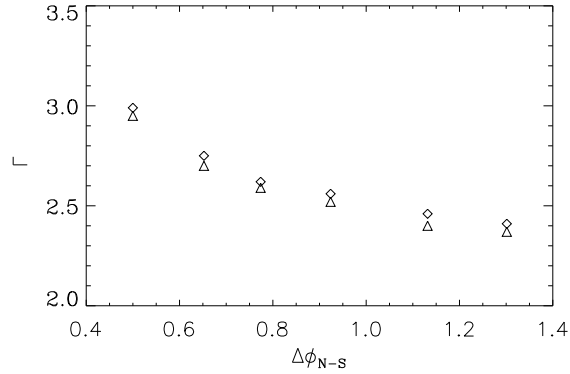}

\caption{Photon index versus $\Delta \phi$ for
$B = 10^{14}~G$ (triangles) and $B = 10^{15}~G$ (diamonds). See text for
details
\label{fgamma}
}
\end{figure*}

\begin{figure*}
\includegraphics[width=84mm]{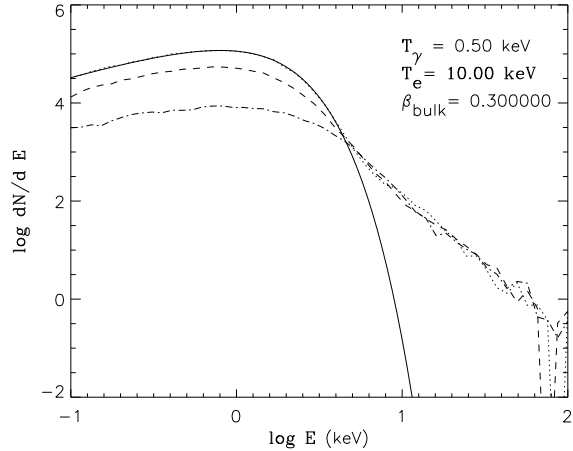}

\caption{
Spectrum from a single emitting patch on the star surface. The LOS is at
$\Theta_s=90^\circ$ and
$\Phi_s=20^\circ$ (dotted line), $140^\circ$ (dashed line) and $220^\circ$
(dash-dotted line). The solid line represents the seed
blackbody. Because photons are collected in a single patch on the sky, the
counting statistics is low at the higher energies and the spectrum looks
``noisy''.
\label{onepatch}
}
\end{figure*}

\begin{figure*}
\hbox{
\includegraphics[width=84mm]{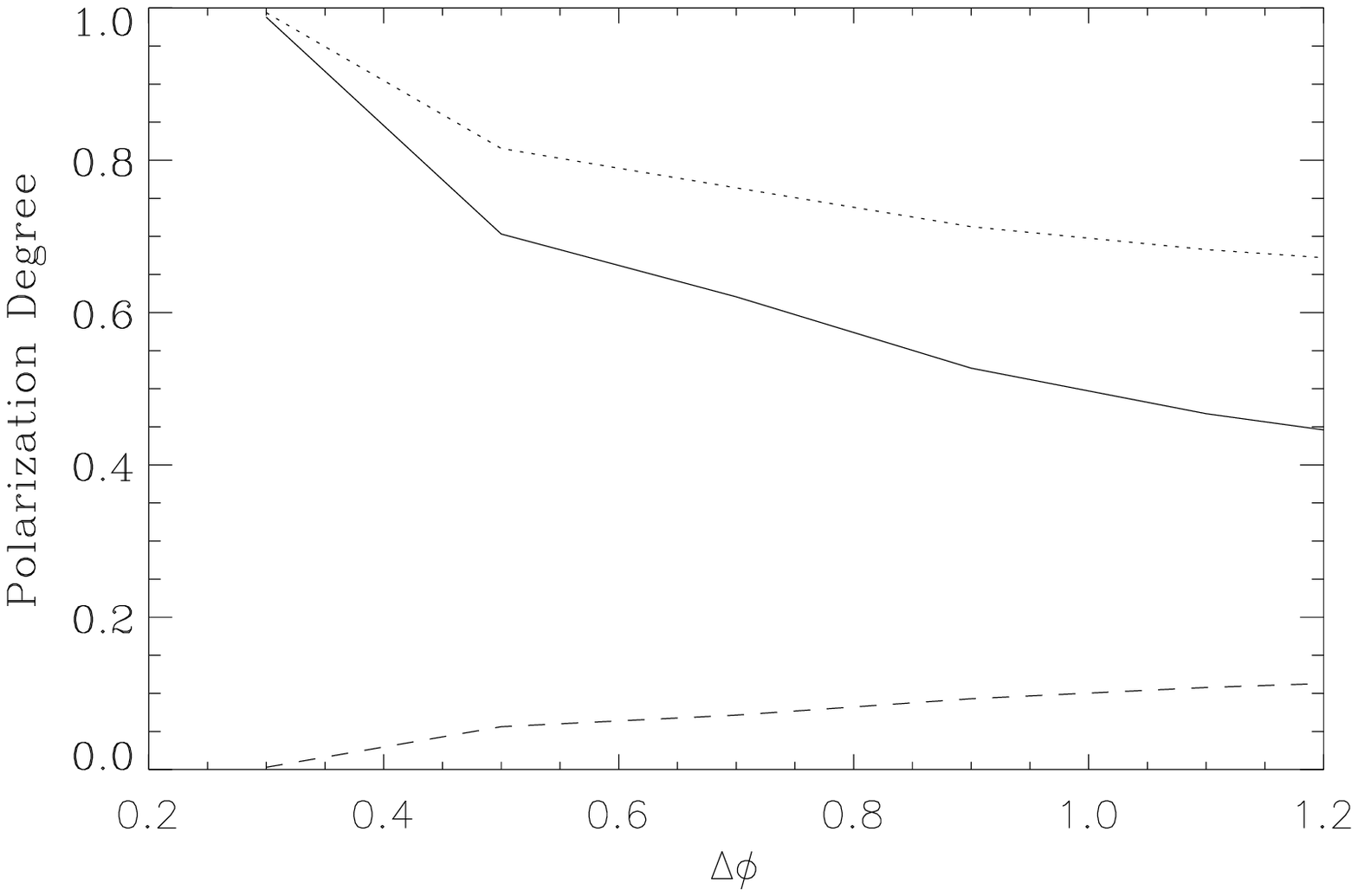}
\includegraphics[width=84mm]{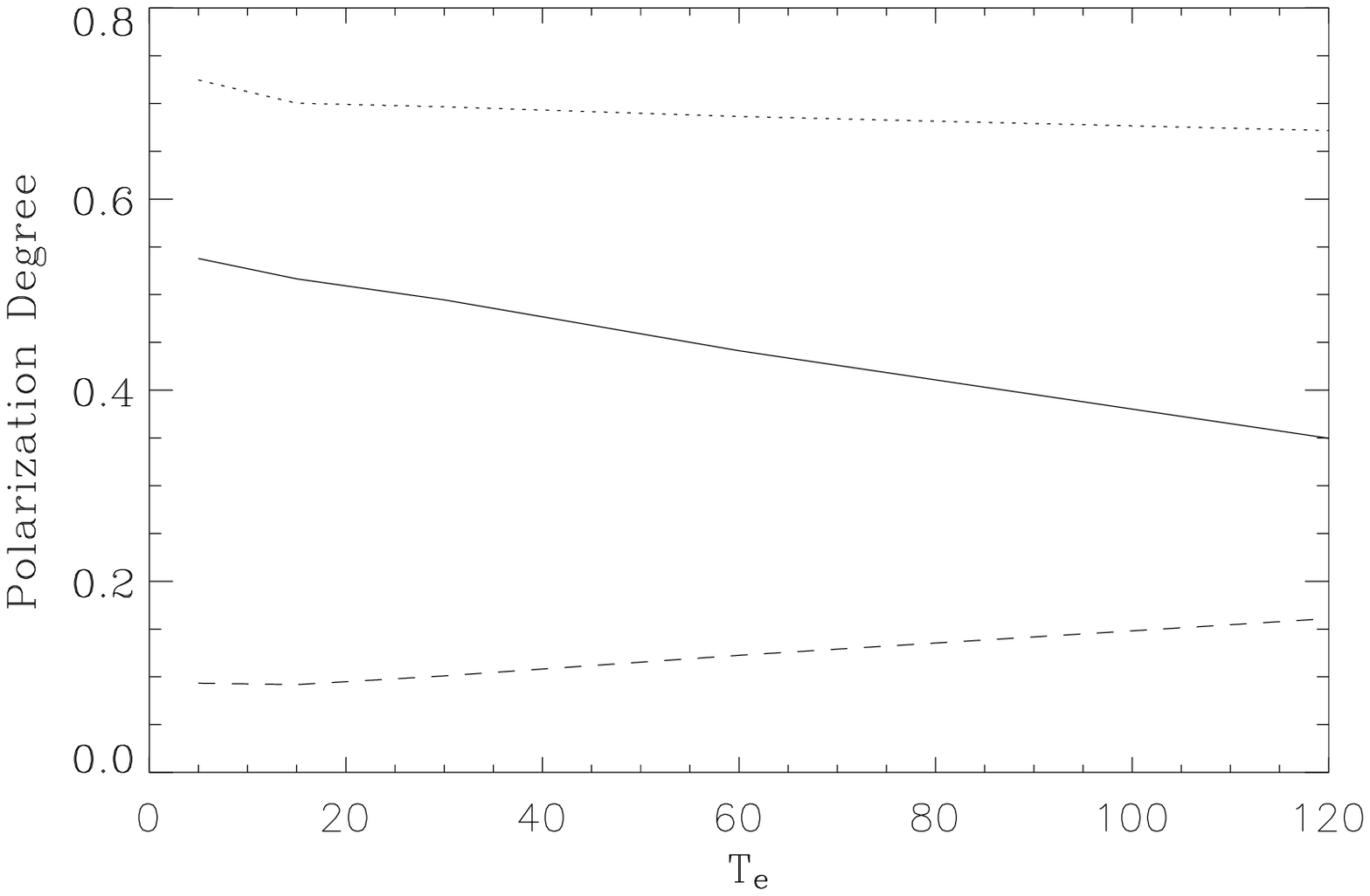}
}

\caption{Left: degree of polarization as a function of $\Delta \phi$ for
$B=10^{14}$~G, $kT = 0.5$~keV, $\beta_{bulk} = 0.3$ and $kT_e = 30$~keV.
Right: Same, but as a function of $kT_e$ for $\Delta \phi = 1$.
In both panels different curves correspond to: seed photons 100\% polarized
in the ordinary (solid line),  extraordinary mode (dotted line), and
unpolarized (dashed line). See text for details.
\label{polla1}
}
\end{figure*}

\begin{figure*}
\includegraphics[width=84mm]{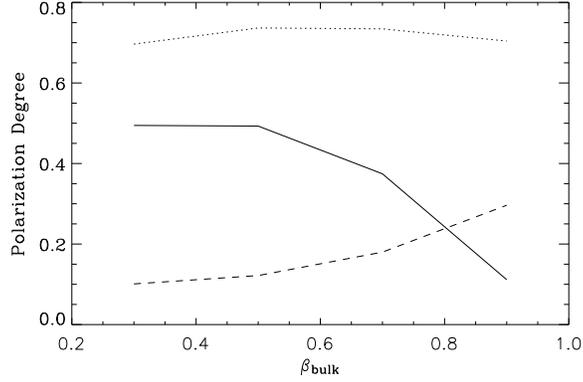}

\caption{Degree of polarization as a function of $\beta_{bulk}$ for
$kT_e = 30$~keV and $\Delta \phi = 1$. Other parameters and line code as in fig. \ref{polla1}.
\label{polla2}
}
\end{figure*}

\begin{figure*}
\hbox{
\includegraphics[width=84mm]{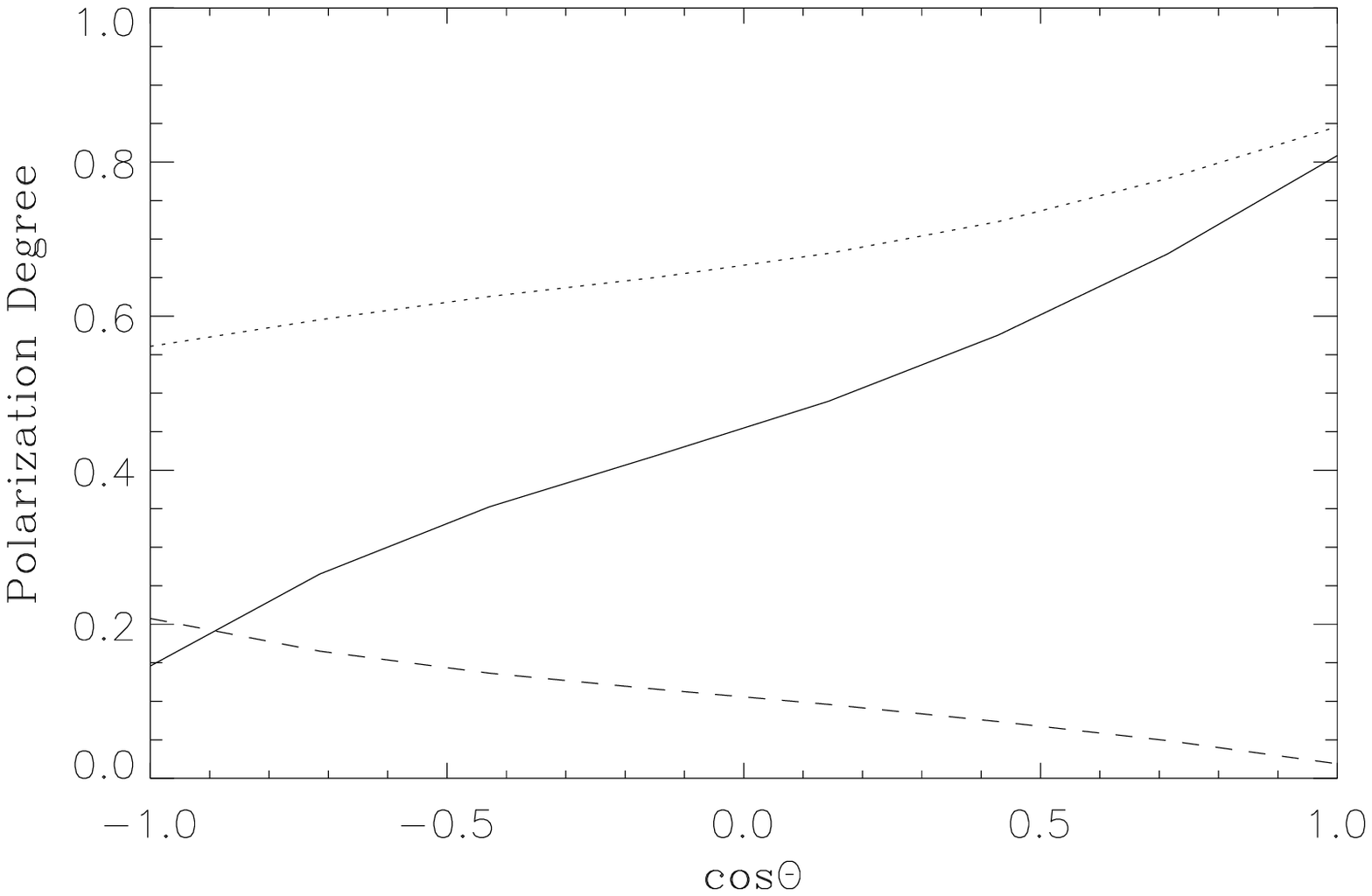}
\includegraphics[width=84mm]{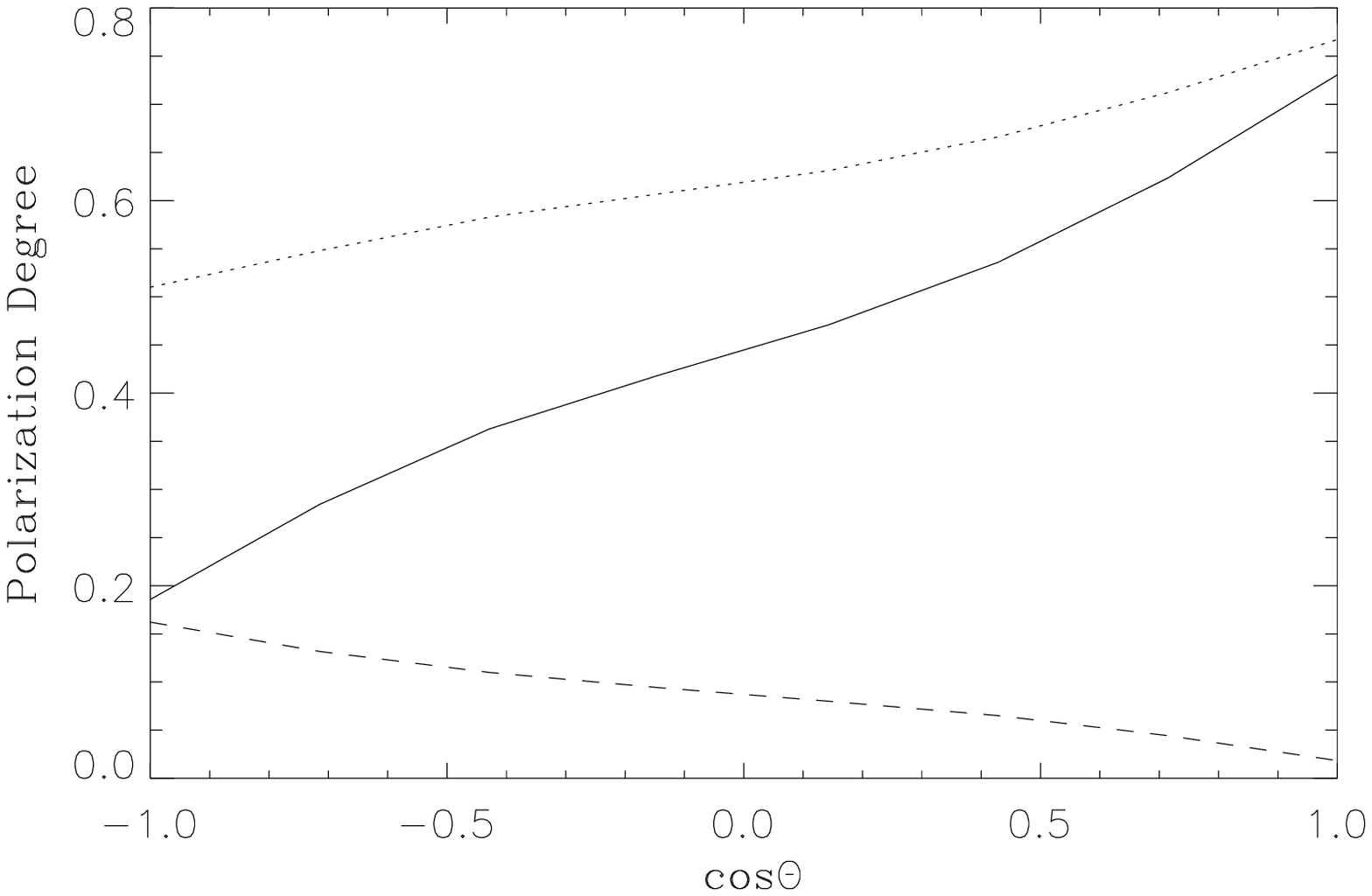}
}
\caption{
Right:
Degree of polarization as a function of the magnetic colatitude $\Theta$
at which seed photons are emitted. Here
$kT = 0.5$~keV, $\beta_{bulk} = 0.3$, $keT_e = 30$~keV and $\Delta \phi =
1$. Left panel:
$B=10^{14}$~G; right panel: $B=10^{15}$~G.
In both panels different curves correspond to: ordinary
(solid line),  extraordinary seed
photons (dotted line) an
unpolarized seed photons (dashed line).
Photons have been
integrated over the entire sky at infinity and over the azimuthal angle at
the star surface.
\label{polla3}
}
\end{figure*}

\begin{figure*}
\hbox{
\includegraphics[width=84mm]{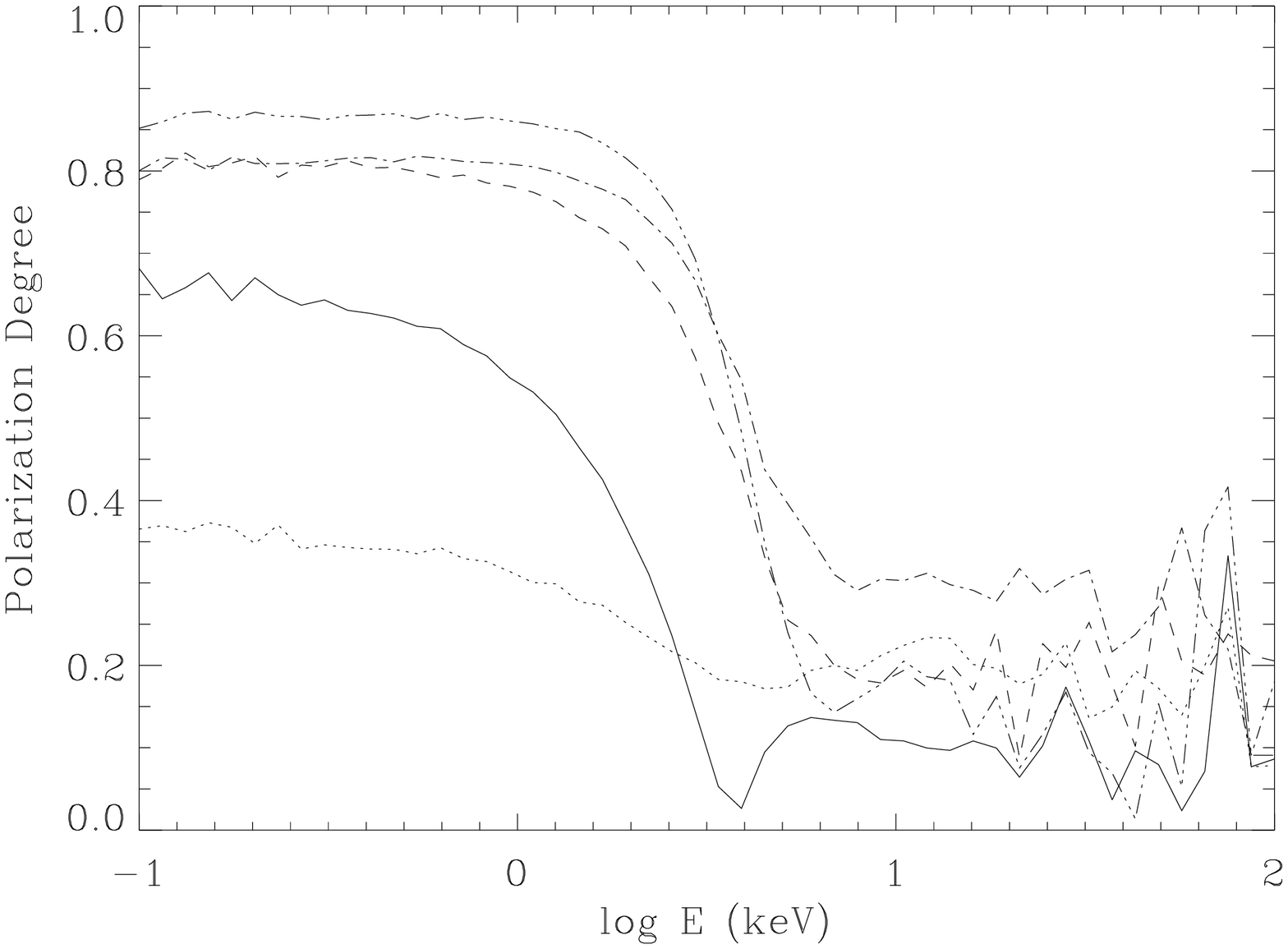}
\includegraphics[width=84mm]{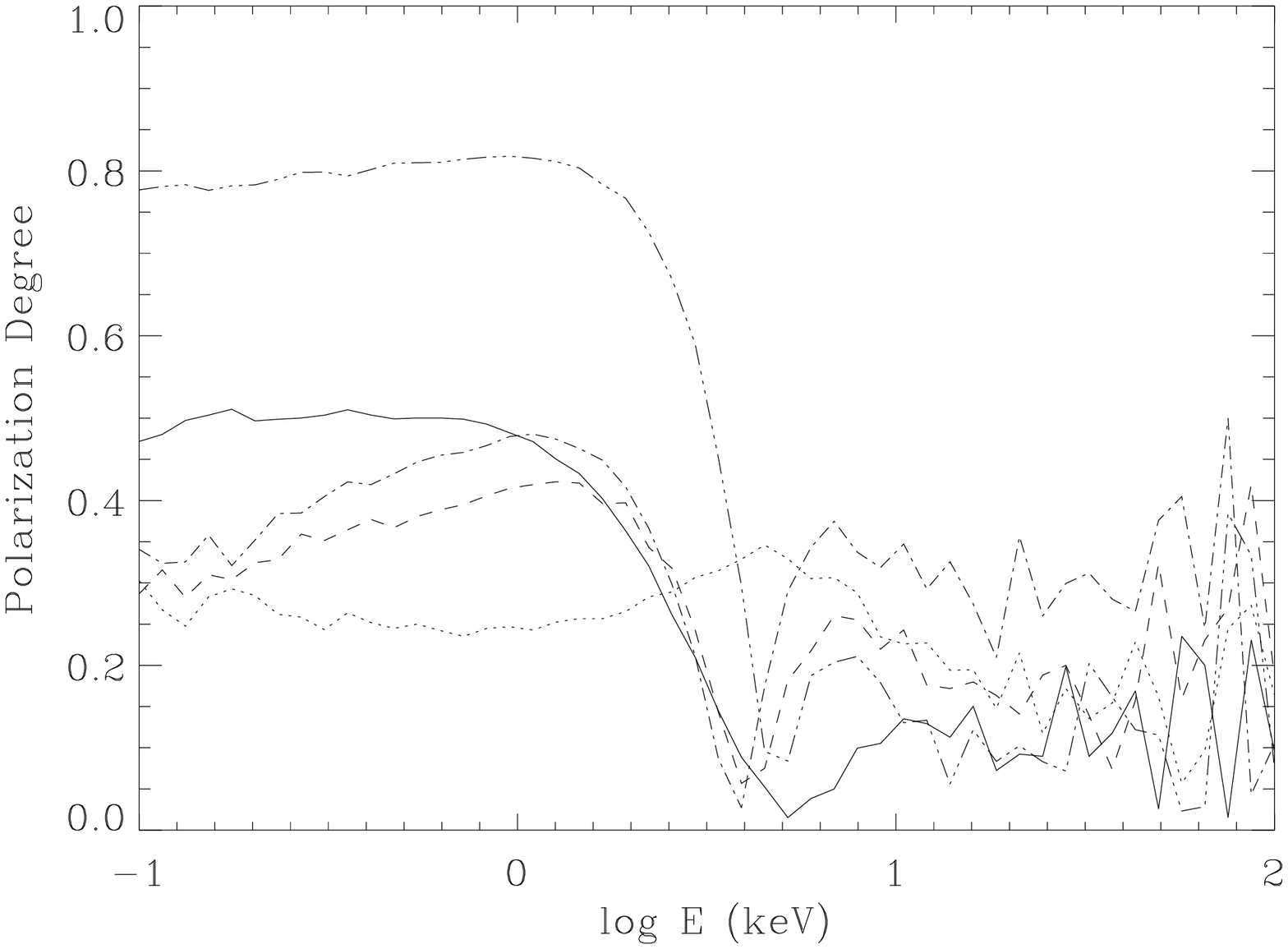}
}
\caption{
Degree of polarization as a function of energy for different values of the
LOS inclination, $\Theta_s=162^\circ$
(solid), $126^\circ$ (dotted), $90^\circ$ (dashed), $54^\circ$ (dash-dotted) and $18^\circ$
(dash-triple dotted).
Left: 100\% extraordinary polarized seed photons.  Right: 100\% ordinary
polarized seed photons. As in Fig. \ref{onepatch}, the low statistics is responsible for the
noisy appearance of the plot at higher energies. The decrease in the polarization degree with
energy is clearly visible notwithstanding.
\label{fpolfreq}
}
\end{figure*}

\begin{figure*}
\includegraphics[width=84mm]{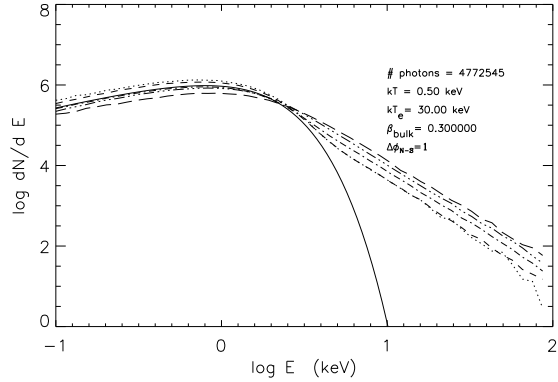}

\caption{ Computed spectra for $B =
10^{14}~G$, $kT  = 0.5$~keV, $kT_e = 30$~keV, $\beta_{bulk} = 0.3$, $\Delta \phi = 1$, $\xi=45^\circ$ and five
different values of the viewing angle $\chi$: $ 0.01^\circ$ (dotted),
$45^\circ$ (short dashed), $90^\circ$ (dash-dotted), $135^\circ$
(dashed-triple dotted), and $180^\circ$ (long dashed). The solid line represents the seed
blackbody. Here seed photons are assumed to be completely polarized in
extraordinary mode. \label{variachi} }
\end{figure*}

\begin{figure*}
\hbox{
\includegraphics[width=84mm]{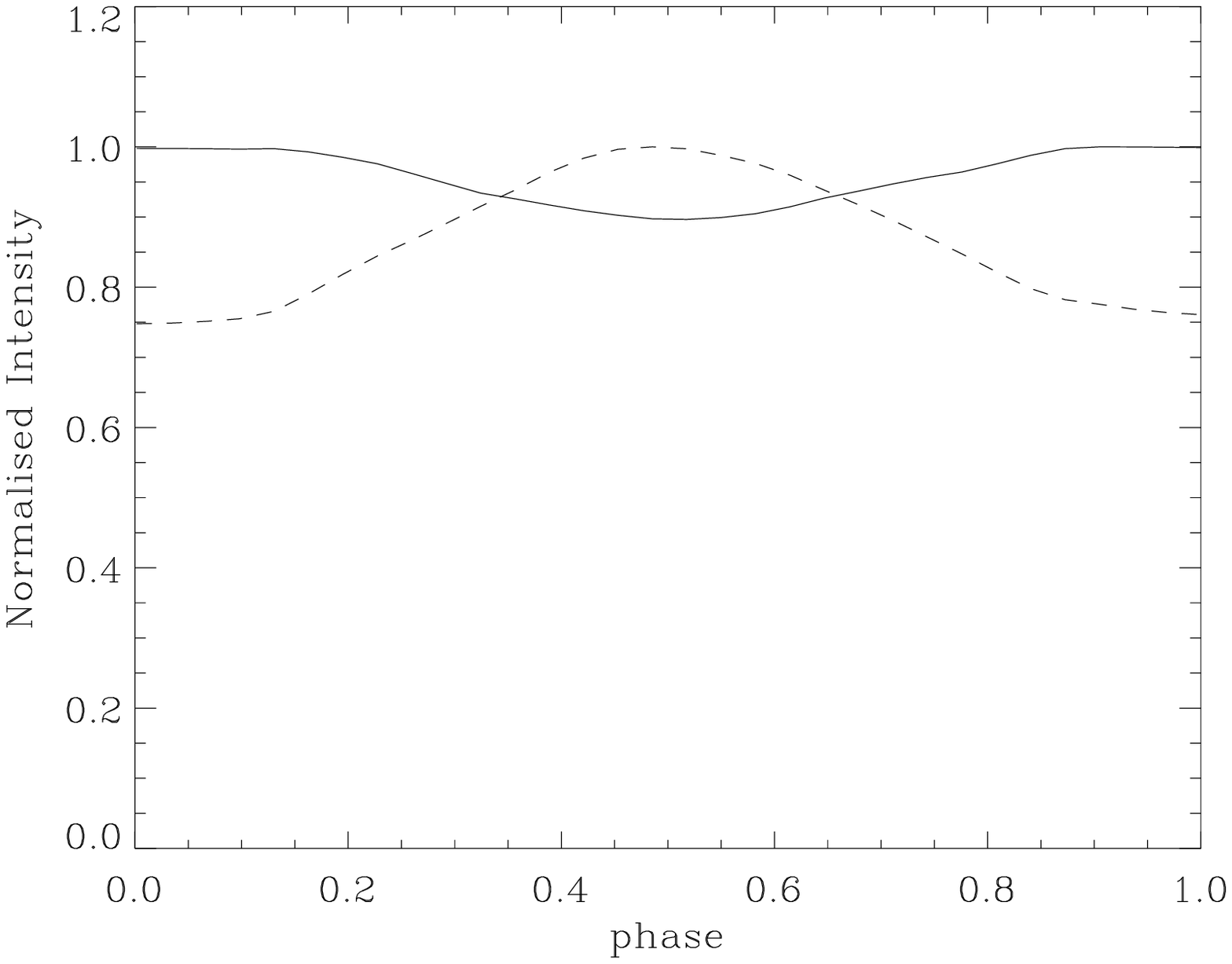}
\includegraphics[width=84mm]{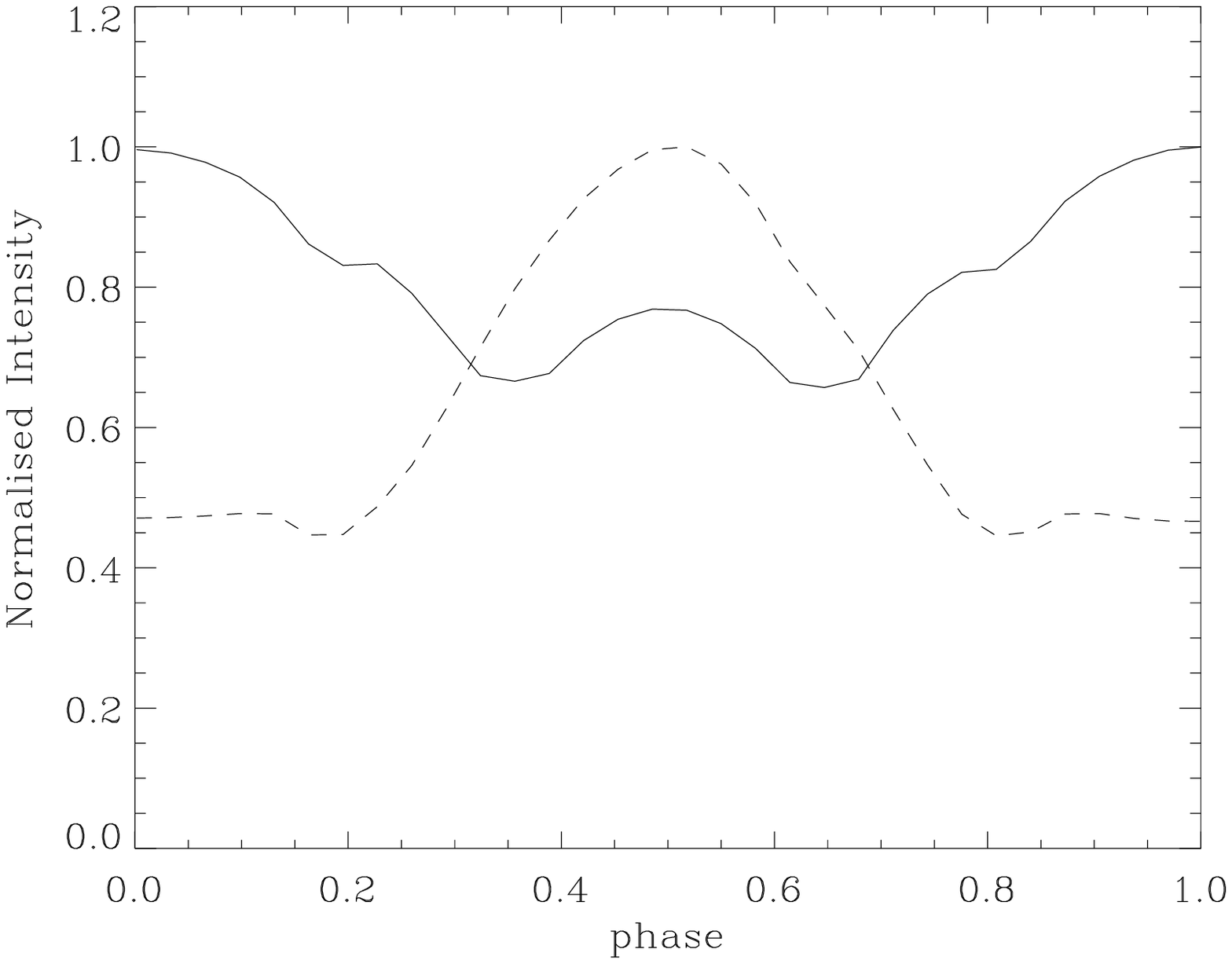}
}
\caption{
Left: the lightcurves in the soft (0.5--2 keV, solid line) and hard X-ray band (2--6 keV, dashed
line); the model parameters are $\chi = 90^\circ$, $\xi = 10^\circ$,
$\Delta\phi = 0.7$, $\beta_{bulk} = 0.3$ and $kT = 0.3\ {\rm keV}$. Right: same as in the left panel, but
for $\xi =50^\circ$. See text for details.
\label{figlc}
}
\end{figure*}

\begin{figure*}
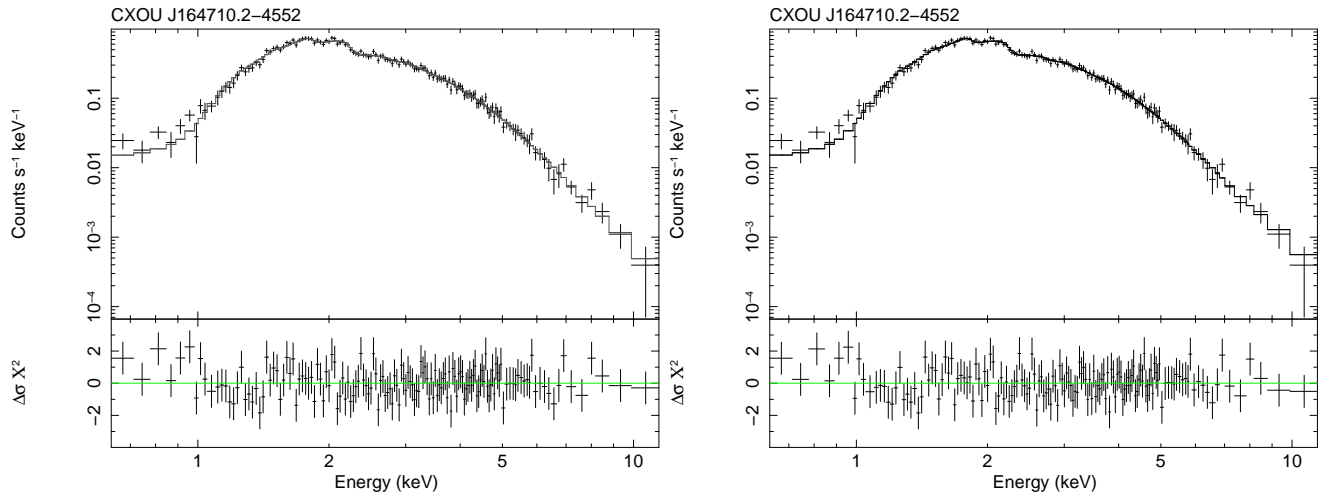

\hbox{
\includegraphics[width=64mm,angle=-90]{fig14a.ps}
\includegraphics[width=64mm,angle=-90]{fig14b.ps}
}
\caption{Left: Fit of the {\it XMM-Newton\/} EPIC-pn spectrum
of \wes\ with an absorbed {\tt ntznoang} model. Top: data and best fit
model; bottom: residuals. Right: the same observation fitted with an absorbed
{\tt ntzang} model.
\label{noang}
}
\end{figure*}

We thank G.L.~Israel for carrying out the preliminary fit of the spectrum
of \wes\ and kindly providing us with Fig.~\ref{noang}, and A.~Albano for
producing the lightcurves shown in Fig.~\ref{figlc}. We are also grateful
to N.~Sartore, who participated in the early stages of this work in partial fulfillment of
the requirements for his undergraduate degree and computed the models used in Fig. \ref{fgamma}. LN and RT are
partially supported by INAF-ASI through grant AAE TH-058. SZ acknowledges STFC for support through an Advanced
Fellowship.

\newpage
\begin{table}
\caption{Best fit values of the spectral parameters.}
 \label{tab1}
 \begin{tabular}{@{}lcc}
  \hline
   Parameters & {\tt ntznoang} & {\tt ntzang}\\
  \hline
  $N_{H}$ &  1.76$^{+0.06}_{-0.05}$ &  1.76$^{+0.04}_{-0.01}$ \\
  $kT$ &  0.625$^{+0.007}_{-0.008}$ & 0.63 $^{+0.07}_{-0.01}$\\
  $\beta_{bulk}$ & 0.60 $^{+0.03}_{-0.02}$ & 0.65  $^{+0.26}_{-0.07}$ \\
  $\Delta \phi$ & 0.40$^{+0.03}_{-0.32}$ &  0.47 $^{+0.03}_{-0.06}$ \\
  $\chi$ & -- & $2.1 \pm 1.8$   \\
  $\xi$ & -- & $82^{+89}_{-56}$ \\
  Norm &  0.081$^{+0.003}_{-0.003}$ &  0.003$^{+0.140}_{-0.000}$ \\
& \\
Flux  & 6  &   6 \\
& \\
$\chi^2_{\nu}$ (dof) & 0.81 (145) &   0.83 (143)\\
  \hline
 \end{tabular}

 \medskip
Errors in the parameters are at 1$\sigma$ confidence level,
$N_{H}$ is in units of $10^{22}\, {\rm cm}^{-2}$, $kT$ is in keV, $\chi, \xi$ are in
degrees and the observed flux (1--10 keV) is in units of $10^{-12}\ {\rm erg\,s\,cm}^{-2}$.

\end{table}

\newpage

\bsp

\label{lastpage}

\end{document}